\documentclass{aa}  

\usepackage{graphicx}
\usepackage{color}
\usepackage{footnote}
\makesavenoteenv{tabular}
\usepackage{wasysym}
\usepackage{amsmath}
\usepackage{tikz}
\usetikzlibrary{shapes.geometric, arrows} 

\usepackage{newtxtext} 
\usepackage{newtxmath} 
\usepackage{orcidlink} 
\usepackage{upgreek}

\usepackage{hyperref}
\hypersetup{
    colorlinks,
    linkcolor={red!50!black},
    citecolor={blue!50!black},
    urlcolor={blue!80!black},
    pdftitle={Time-dependent Monte Carlo continuum radiative transfer},
    pdfauthor={A. Bensberg and S. Wolf},
    pdfkeywords={radiative transfer, methods: numerical,  circumstellar matter, protoplanetary disks, stars: variables,  ISM: clouds
    }
}

\tikzstyle{start} = [rectangle, rounded corners, minimum width=3cm,
                      minimum height=1cm,text centered, draw=black,
                      fill=orange!30]
\tikzstyle{tstep} = [rectangle, rounded corners, minimum width=3cm,
                          minimum height=1cm,text centered, text width=5cm,
                          fill=blue!30]
\tikzstyle{tstep1} = [rectangle, rounded corners, minimum width=3cm,
                          minimum height=1cm,text centered, text width=5cm,
                          fill=orange!30]
\tikzstyle{tstep2} = [rectangle, rounded corners, minimum width=3cm,
                          minimum height=1cm,text centered, text width=5cm,
                          fill=green!30]
\tikzstyle{tstep3} = [rectangle, rounded corners, minimum width=3cm,
                          minimum height=1cm,text centered, text width=5cm,
                          fill=red!30]
\tikzstyle{io} = [trapezium, trapezium left angle=70, trapezium right angle=110, minimum width=3cm,
                        minimum height=1cm, text centered, text width=4cm,
                          fill=yellow!50]
\tikzstyle{arrow} = [thick,->,>=stealth] 

\begin{document} 

    \title{Time-dependent Monte Carlo continuum radiative transfer}
    
    \author{A. Bensberg \orcidlink{0000-0001-6789-0296} \and S. Wolf \orcidlink{0000-0001-7841-3452}}
    
    \institute{Institut für Theoretische Physik und Astrophysik, Christian-Albrechts-Universität zu Kiel, Leibnizstr. 15, 24118 Kiel, Germany\\
    \email{\href{mailto:abensberg@astrophysik.uni-kiel.de}{abensberg@astrophysik.uni-kiel.de}}}
    
    \date{Received / Accepted }
    
    \abstract
    {Variability is a characteristic feature of young stellar objects that is caused by various underlying physical processes. Multi-epoch observations in the optical and infrared combined with radiative transfer simulations are key to study these processes in detail.}
    {We present an implementation of an algorithm for 3D time-dependent Monte Carlo radiative transfer. It allows one to simulate temperature distributions as well as images and spectral energy distributions of the scattered light and thermal reemission radiation for variable illuminating and heating sources embedded in dust distributions, such as circumstellar disks and dust shells on time scales up to weeks.}
    {We extended the publicly available 3D Monte Carlo radiative transfer code \texttt{POLARIS} with efficient methods for the simulation of temperature distributions, scattering, and thermal reemission of dust distributions illuminated by temporally variable radiation sources. The influence of the chosen temporal step width and the number of photon packages per time step as key parameters for a given configuration is shown by simulating the temperature distribution in a spherical envelope around an embedded central star. The effect of the optical depth on the temperature simulation is discussed for the spherical envelope as well as for a model of a circumstellar disk with an embedded star. Finally, we present simulations of an outburst of a star surrounded by a circumstellar disk.}
    {The presented algorithm for time-dependent 3D continuum Monte Carlo radiative transfer is a valuable basis for preparatory studies as well as for the analysis of continuum observations of the dusty environment around variable sources, such as accreting young stellar objects. In particular, the combined study of light echos in the optical and near-infrared wavelength range and the corresponding time-dependent thermal reemission observables of variable, for example outbursting sources, becomes possible on all involved spatial scales.}
    {}
    
    \keywords{radiative transfer -- methods: numerical --
    circumstellar matter -- protoplanetary disks -- stars: variables: general -- ISM: clouds}
    
    \maketitle

\section{Introduction}  \label{sec:intro}
Variability is known as a characteristic feature of young stellar objects (YSOs) since their first observations \citep{Joy1945}. The study of these time-dependent phenomena can reveal different underlying physical mechanisms such as spots on the stellar surface caused by magnetic fields, variable mass accretion onto the star, and obscuration by circumstellar material along the line of sight to the observer \citep{Herbst1994}. Since all of these mechanisms have an impact on the measured flux of the YSO, observed light curves of variable YSOs show different shapes ranging from periodical to linear or curved \citep{Park2021}. The time scales of the underlying event can be on the order of hours to years (e.g.,~\citealp{Grankin2007}, \citealp{Cody2014}, \citealp{Gunther2014}). in the case of the extreme accretion events that can be found in FUor and EXor objects, the outburst can even last decades (e.g.,~\citealp{Hartmann1997}, \citealp{Herbig2008}). The variability of the central object does also affect the illumination and thus the heating of the dust of the circumstellar disk. Therefore, variability can also be studied in the scattered light and traced in the thermal reemission radiation of the surrounding dust (e.g.,~ \citealp{Carpenter2001}, \citealp{Johnstone2013}).

Radiative transfer simulations play an important role in the interpretation of astronomical observations. Because of its flexibility, the Monte Carlo method is often applied to simulate the radiative transfer in different astrophysical objects (e.g., \citealp{Wolf1999}; \citealp{Harries2000}; \citealp{Pinte2006}; \citealp{Dullemond2012}; \citealp{Ober2015}). However, an underlying assumption in most of these solutions is that of radiative equilibrium. Moreover, while there is an implementation of Monte Carlo radiative transfer that uses gas properties to approach time-dependent heating and cooling (\citealp{Harries2011}), there is currently no solution available focusing on the thermal properties of the dust.

In this paper, we present an implementation of an algorithm for full time-dependent 3D Monte Carlo radiative transfer (MCRT), which includes simulations of the scattered light, dust thermal reemission radiation, and temperature distribution with a focus on the latter. For this purpose, we extend the publicly available 3D Monte Carlo radiative transfer code \texttt{POLARIS}\footnote{\url{https://portia.astrophysik.uni-kiel.de/polaris/}} \citep{Reissl2016}, which is used for the simulation of various astrophysical objects ranging from galaxies (\citealp{Pellegrini2020}, \citealp{Reissl2020}) and Bok globules (\citealp{Brauer2016}, \citealp{Zielinski2021}) to circumstellar disks (\citealp{Brauer2019}, \citealp{Brunngraber2020}) and exoplanetary atmospheres (\citealp{Lietzow2021}).

This paper is organized as follows: The computational method is described in Sect.~\ref{sec:method}. In Sect.~\ref{subsec:phys} we focus on the theoretical background of the relevant thermal processes. The basic algorithm for the time-dependent simulation of temperature distributions is presented in Sect.~\ref{subsec:timedep_temp}. The procedures for the calculation of high-resolution images of the thermal reemission radiation and scattered light, that is to say algorithms for time-dependent ray tracing as well as time-dependent scattering, are briefly described in Sect.~\ref{subsec:timedep_ray} and~\ref{subsec:timedep_sca}. An overview of different tests of the implemented routines is given in Sect.~\ref{sec:test}. The impact of the temporal step width and the number of photon packages per time step is discussed in Sect.~\ref{subsec:dt} and \ref{subsec:nph}. The influence of the optical depth on the temperature distribution simulations for 1D and 2D models are examined in Sect.~\ref{subsec:1D_depth} and \ref{subsec:2D_depth}, respectively. In Sect.~\ref{subsec:2D_burst}, we discuss the impact of a luminosity outburst of a central star embedded in a circumstellar dust disk on the resulting dust temperature distribution, as well as on the scattered light and thermal reemission radiation. Finally, a summary as well as an outlook on further applications and potential improvements of the method are provided in Sect.~\ref{sec:disc}.

\section{Computational method}  \label{sec:method}
In the following, we outline the theoretical background of time-dependent continuum radiative transfer. After discussing the thermal processes and properties of the dust in Sect.~\ref{subsec:phys}, we describe the implementation of a time-dependent method of radiative transfer simulations for temperature, dust reemission and scattering in Sect.~\ref{subsec:timedep_temp}. For this, we extend the publicly available 3D Monte Carlo radiative transfer code \texttt{POLARIS} \citep{Reissl2016}.

\subsection{Description of heating and cooling} \label{subsec:phys}
\begin{figure}[!ht]
    \centering
    \includegraphics[width=1.\linewidth]{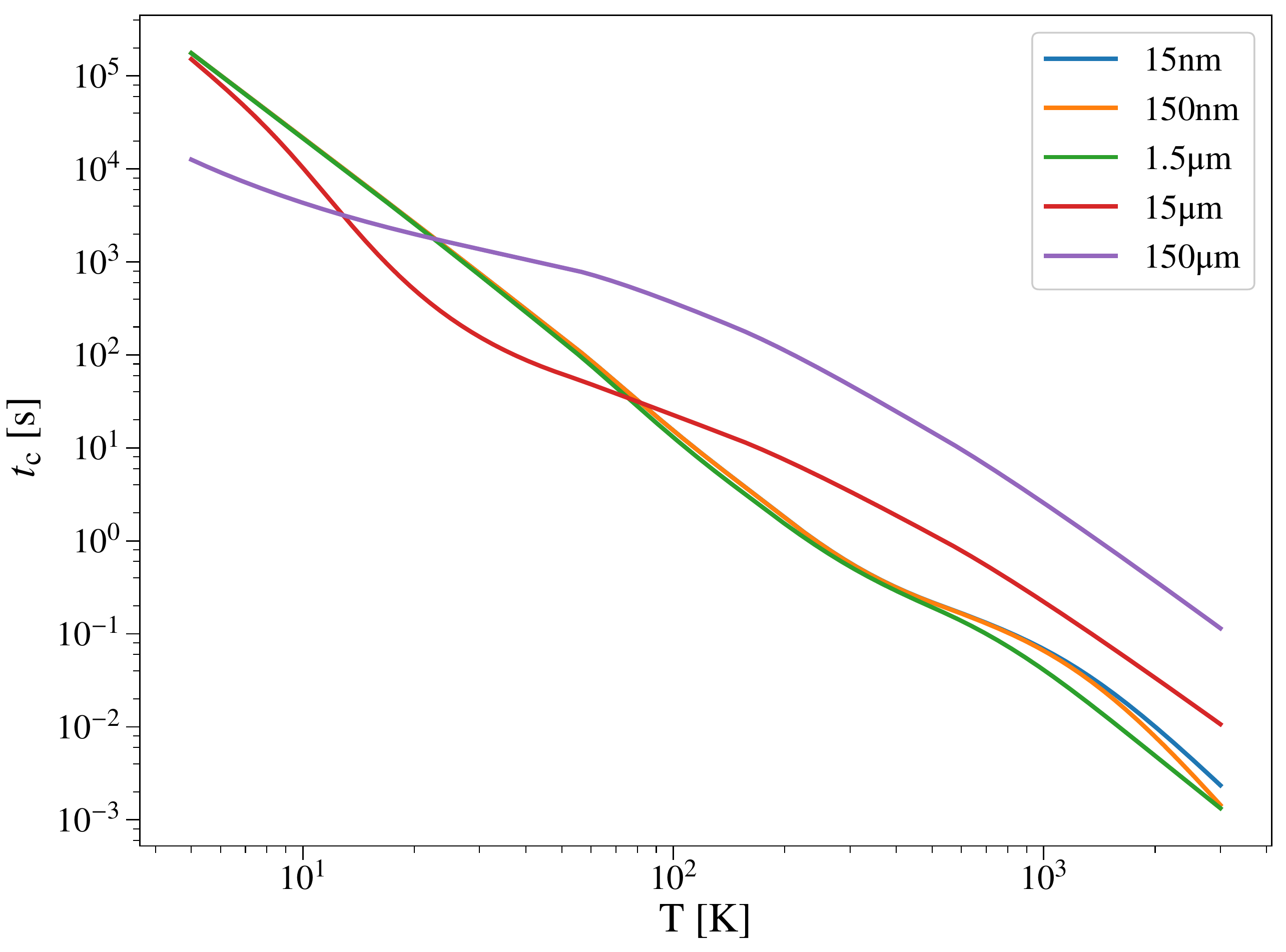}
    \caption{Cooling time described in Eq.~\ref{eq:cool} for different temperatures of grains with different radii consisting of astronomical silicate.}
    \label{fig:cool}
\end{figure}

We first focus on the thermal processes and corresponding properties of the dust. 
The $i$-th photon package with wavelength $\lambda$ emitted by a radiation source with luminosity $L$ that is emitting $N_{\rm ph}$ photon packages in the time-interval $\Delta t$ carries the energy
\begin{equation}\label{eq:pp_en}
    e_{i,\lambda} = \frac{L \Delta t}{N_{\rm ph}}.
\end{equation}
Along its path through the model space, the dust absorbs energy of the photon package in relation to the wavelength-dependent mass absorption coefficient $\kappa_{\lambda}$. The total absorption rate $\dot{A}$ in a region with mean radiation intensity $J_{\lambda}$ is then given by:
\begin{equation}\label{eq:abs}
    \dot{A} = 4\pi \int_0^\infty \kappa_{\lambda} J_{\lambda} \textrm{d}\lambda.
\end{equation}
Following \cite{Lucy1999}, the mean intensity $J_{\lambda}$ of a discrete number of photon packages with energy $e_{i,\lambda}$ within a wavelength range d$\lambda$ traveling a path length $\delta l_i$ in a volume $V$ in the time-interval $\Delta t$ can be estimated by:
\begin{equation}\label{eq:lucy}
    J_{\lambda}\textrm{d}\lambda = \frac{1}{4\pi} \frac{1}{V \Delta t} \sum_i e_{i,\lambda} \delta l_i.
\end{equation}
An estimator for the absorption rate in Eq.~\ref{eq:abs} is then derived by substitution of Eq.~\ref{eq:lucy} and discretization:
\begin{equation}\label{eq:abs_est}
    \dot{A} = \frac{1}{V \Delta t} \sum_i \kappa_{\lambda} e_{i,\lambda} \delta l_i.
\end{equation}
Due to the absorption process the dust will heat up to a temperature $T$, that is the enthalpy $u(T)$ of the dust grain will increase. According to \cite{Guhathakurta1989} the enthalpy $u(T)$ of an $N$ atom dust grain with volume $V_{\rm d}$ and heat capacity per volume $C_{\rm v}$ can be calculated via:
\begin{equation}\label{eq:enth}
    u(T) = (1-2/N)V_{\rm d} \int_0^T C_{\rm v} (T) \textrm{d}T.
\end{equation}
Under the assumption of local thermodynamic equilibrium (LTE), the dust will emit radiation with an emission rate $\dot{E}$ corresponding to its emission spectrum $B_{\lambda}$ at temperature $T$ scaled with the absorption coefficient $\kappa_{\lambda}$:
\begin{equation}\label{eq:emi}
    \dot{E}(T) = 4\pi \int_0^\infty \kappa_{\lambda} B_{\lambda}(T) \textrm{d}\lambda.
\end{equation}
Considering the difference between absorption and emission rate within one time step, that is combining Eqs.~\ref{eq:abs_est} and~\ref{eq:emi}, we now can estimate the enthalpy $u_{n+1}$ of the dust after $n+1$ time steps with step width $\Delta t$:
\begin{equation}\label{eq:nextu}
    u_{n+1} = u_n + (\dot{A} - \dot{E}) \Delta t.
\end{equation}
However, this holds only true if we assume the absorption and emission rate is constant over the step width $\Delta t$. We can constrain the temporal step width by considering the time it takes to emit the total enthalpy of a dust grain at temperature $T$, that is the cooling time $t_{\rm c}(T)$. For the case of emission without absorption, we find
\begin{equation}\label{eq:cool}
    t_{\rm c}(T) = \frac{u(T)}{\dot{E}(T)} 
\end{equation}
as upper limit for the step width. Since both emission rate and enthalpy depend on the size of the grain, the cooling time is size-dependent. An exemplary plot of the cooling time of astronomical silicate (see, Sect.~\ref{sec:test}) with different grain sizes is shown in Fig.~\ref{fig:cool}. 
It can be seen that the cooling times of grains with sizes up to $1.5\;\upmu$m are of the same order of magnitude. Larger grains are cooling slower, because they are less efficient emitters due to a smaller ratio of their emitting surface to their volume. In both cases, the cooling time is rapidly decreasing with increasing temperature. The choice of the temporal step width $\Delta t$ is thus crucial for the goodness of the estimators described in Eqs.~\ref{eq:abs_est} and~\ref{eq:nextu}. A detailed discussion of the corresponding numerical consequences can be found in Sect.~\ref{sec:exam}.

\subsection{Numerical implementation}\label{subsec:timedep_temp}
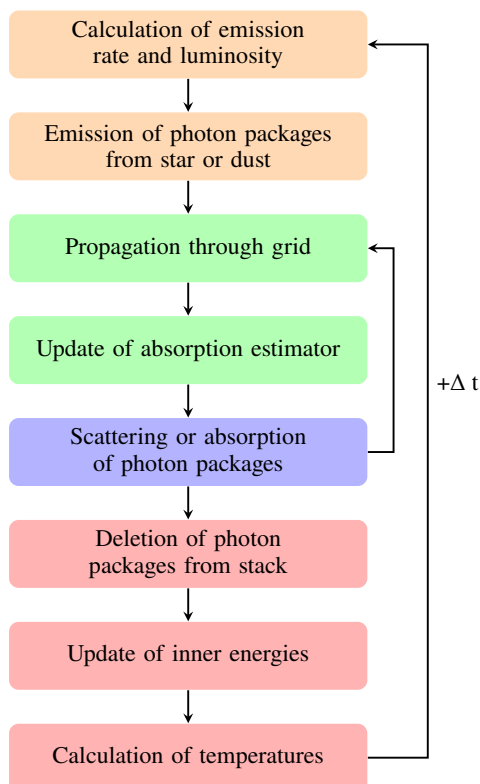
\begin{figure}
    \centering
    \resizebox{0.35\textwidth}{!}{
        \begin{tikzpicture}[node distance=1.5cm] 
            \node (cell_lum) [tstep1] {Calculation of emission\\ rate and luminosity};
            \node (emission) [tstep1, below of=cell_lum] {Emission of photon packages\\ from star or dust};
            \node (prop) [tstep2, below of=emission] {Propagation through grid};
            \node (abs_est) [tstep2, below of=prop] {Update of absorption estimator};
            \node (sca_abs) [tstep, below of=abs_est] {Scattering or absorption\\ of photon packages};
            \node (delete) [tstep3, below of=sca_abs] {Deletion of photon packages from stack};
            \node (enth) [tstep3, below of=delete] {Update of inner energies};
            \node (temps) [tstep3, below of=enth] {Calculation of temperatures};
            
            \draw [arrow] (cell_lum) -- (emission);
            \draw [arrow] (emission) -- (prop);
            \draw [arrow] (prop) -- (abs_est);
            \draw [arrow] (abs_est) -- (sca_abs);
            \draw [arrow] (sca_abs) -- ++ (3cm,0) |- (prop);
            \draw [arrow] (sca_abs) -- (delete);
            \draw [arrow] (delete) -- (enth);
            \draw [arrow] (enth) -- (temps);
            \draw [arrow] (temps) -- ++ (3.5cm,0) |- node[below right, pos=0.275]{+$\Delta$ t} (cell_lum);
        \end{tikzpicture}
    }
    \caption{Flowchart of the numerical procedure for the simulation of time-dependent temperature distributions described in Sect.~\ref{subsec:timedep_temp} (the four main steps outlined there are color coded: 1-orange, 2-green, 3-blue and 4-red).}
    \label{fig:flowchart}
\end{figure}

Based on the concepts described above, we implemented a time-dependent continuum radiative transfer algorithm in our 3D MCRT code \texttt{POLARIS}. Therefore, the temporal step width is added as parameter to the stationary radiative transfer scheme for the simulation of the temperature distribution (hereafter: stationary method). A detailed description of the stationary scheme can be found for example in \cite{Wolf1999} and \cite{Reissl2016}. 
\texttt{POLARIS} allows the use of several grid geometries to build the model space including cylindrical, spherical, octree, and voronoi. The number of grid cells as well as their spacing can be set individually for every dimension. Multiple stellar sources can be set in the grid at any position. However, for the sake of simplicity we are considering the case of a single central star (spherical, isotropic irradiation) for the following description. A schematic flowchart of the whole procedure for a single time step is shown in Fig.~\ref{fig:flowchart}. 
The algorithm for the simulation of the time-dependent temperature distribution can be divided into a series of four steps that are repeated every time step until the total simulation time, that is the time at which the simulation is stopped, is reached.

\paragraph{1. Emission from dust or star}
At the beginning of each time step, a fixed number of photon packages $N_{\rm ph}$ is emitted from the star or the dusty environment. Therefore, we have to determine the luminosity of the dust of every cell. This is done by reading out the current temperature of the cell from the grid and using Eq.~\ref{eq:emi} to calculate the respective emission rate of a single dust grain. We can now calculate the luminosity of the $i$-th cell by multiplying the emission rate $\dot{E}_i$ of the cell with its number density of the dust grains $n_i$ and volume $V_i$: $L_i = \dot{E}_i V_i n_i$. The total dust luminosity $L_{\rm{tot}}$ is then obtained by summing up the luminosity of all cells. The ratio of total dust luminosity to the combined luminosity of star and dust gives the probability of a photon package to be emitted from the corresponding sources. in the case of thermal emission from within the dust distribution, the emitting cell is determined by the cumulative probability density function $p_i$ for the $i$-th cell given by:
\begin{equation}\label{eq:cell_prob}
    p_i = \left(\sum\limits_{k=0}^{i} L_k\right)/L_{\rm{tot}}.
\end{equation}
Once the source is determined, a random direction of emission is drawn and a wavelength is assigned to the photon package. The respective cumulative probability density function $p_{\lambda_i}$ of a wavelength distribution with bin size $\Delta \lambda$ and a source luminosity $L_{\lambda_i}$ at wavelength $\lambda_i$ is given by:
\begin{equation}\label{eq:wave_prob}
    p_{\lambda_i} = \left(\int_{\lambda_i}^{\lambda_i + \Delta \lambda} L_{\lambda_i}d\lambda\right)/L_i.
\end{equation}
Next the energy of the photon package is assigned using Eq.~\ref{eq:pp_en} and the first point of interaction at optical depth $\tau$ is determined by using a random number $\zeta \in [0,1)$ (see also~\citealp{Reissl2016}):
\begin{equation}\label{eq:optdep}
    \tau = -\log(1-\zeta).
\end{equation}
Lastly, the photon package is stored by adding it to a stack (hereafter: photon stack) that is carried over from one time step to another. The emission process of a photon package can be divided into the following steps:
\begin{enumerate}
    \item Determination of the source of emission by using the probability density function determined by the luminosity ratio of the emitting sources
    \item Calculation of the emitting cell in the case of thermal reemission by dust using Eq.~\ref{eq:cell_prob}
    \item Computation of the random direction of emission
    \item Determination of the wavelength using the cumulative probability density function of the emission spectrum of each source 
    \item Calculation of the energy of the photon packages according to Eq.~\ref{eq:pp_en}
    \item Computation of the optical depth to first interaction via Eq.~\ref{eq:optdep}
    \item Storing of the photon package (photon stack)
\end{enumerate}
This procedure is repeated until all $N_{\rm ph}$ photon packages are emitted.

It should be noted that the emission of photon packages with equally distributed energy and wavelength determined by the cumulative probability density function of the emission spectrum discussed above will overestimate the energy at the most probable wavelength if the number of photons is not high enough to be statistically significant. While this will affect the local radiation field, it will not significantly affect the resulting dust temperature distribution, since the local deviations average out on a global scale. However, to make sure that the local radiation field is reproduced correctly, steps 4 and 5 can be replaced by another emission method: Instead of drawing the wavelength of photon packages using a probabilistic approach, a fixed number of photon packages $N_{\lambda}$ is emitted for every wavelength bin and the energy is assigned using the respective value of the emission spectrum. The energy $e_{\star,j}$ of a photon package with wavelength $\lambda$ for a stellar source with radius $R_\star$ and effective temperature $T$ is then given by:
\begin{equation}\label{eq:alt_pp_en}
e_{\star,j} = 4\pi^2R^{2}_{\star} \frac{B_\lambda(T) \Delta t}{N_{\lambda}}.
\end{equation}
in the case of thermal reemission radiation by the dust, the emitting cells are determined by the luminosity ratio described by the probability density function in Eq.~\ref{eq:cell_prob}. This is done to save computation time by neglecting the thermal reemission of dust with low temperatures that will not significantly contribute to the radiation field. The wavelength distribution of the emitted photon packages depends on the temperature of the dust of the corresponding cell. Since photon packages will only be emitted from cells with high dust luminosity, it must be ensured that the total luminosity $L_{\rm{tot}}$ of the dust of all cells is conserved. To do this for the emission of photon packages of every wavelength of the chosen cells, Eq.~\ref{eq:pp_en} is modified by scaling with the emission spectrum $\tilde{B}_\lambda(T)$ of the respective cell. Here, the emission spectrum is normalized such that the integral over the wavelength is equal to 1. The energy $e_{\textrm{dust}, j}$ of a photon package emitted by the dust is then calculated via:
\begin{equation}\label{eq:alt_pp_en_dust}
e_{\textrm{dust}, j} = \frac{L_{\rm{tot}} \Delta t}{N_{\lambda}} \cdot \tilde{B}_\lambda(T).
\end{equation}
However, this method is computationally expensive because $N_{\lambda}$ photon packages need to be emitted per wavelength, emitting source and time step. Thus, it is only used if the detailed, local radiation field is of interest.
 
\paragraph{2. Continuous absorption}
In the next step, all photon packages of the photon stack are propagated through the grid until either the optical depth of interaction (see, Eq.~\ref{eq:optdep}) or the end of the time step is reached. This is determined by integrating the path length and the optical depth of each photon package. While traveling through a cell, a photon package will deposit energy that adds to the estimated absorption rate of the cell, which is determined using Eq.~\ref{eq:abs_est}. The photon package is not loosing energy on its way through the grid, that is, the absorbed energy is not subtracted. However, there are other approaches to handle absorption that deliver equivalent results, but require higher computational times. A more detailed discussion of these different approaches can be found in Appendix~\ref{sec:abs_est}. 

\paragraph{3. Scattering or absorption}
When the optical depth of interaction determined by Eq.~\ref{eq:optdep} is reached, the probability for scattering and absorption is determined by the albedo of the dust grains. The type of interaction is then obtained by using a random number drawn from a uniform distribution. Since the process of choosing the type of interaction and the treatment of scattering are not different from the stationary case, we refer to~\cite{Reissl2016} for a detailed description. in the case of an absorption event, the photon package is marked for deletion without adding additional energy to the absorbing cell.
If the photon package is scattered, a new optical depth for interaction is calculated using Eq.~\ref{eq:optdep} and the photon package is propagated further through the grid, that is steps 2 and 3 are repeated until the total path length of the photon package reaches the maximum path length corresponding to the temporal step width. Photon packages leaving the grid within the time step will also be marked for deletion.

\paragraph{4. Update of enthalpy and temperature}
At the end of a time step, the photon packages marked for deletion are removed from the stack and the temperature of each cell is updated. Therefore, the new enthalpy $u(T)$ of the dust of each cell is calculated by evaluating Eq.~\ref{eq:nextu}. The corresponding temperature is determined by interpolating precalculated values of the enthalpy using Eq.~\ref{eq:enth}. After the temperatures are updated, the algorithm proceeds with the next time step.

\subsection{Time-dependent ray tracing}\label{subsec:timedep_ray}
The algorithm presented in Sect.~\ref{subsec:timedep_temp} allows producing images of the scattered stellar radiation and the thermal reemission radiation of the dust. However, the simulation of high resolution images using this algorithm is computationally very expensive since every photon package needs to be stored on and loaded from the memory in every time step. Therefore, we implemented two fast, time-dependent methods to calculate scattering images and dust emission maps. For the latter, a time-dependent ray tracing algorithm is used. The time-dependent algorithm for the simulation of images of the scattered light is presented in Sect.~\ref{subsec:timedep_sca}.

The ray tracer implemented in \texttt{POLARIS} follows the path of parallel rays through a model space, which is divided into multiple grid cells. A detector with a two-dimensional array of pixels is set outside of the model space. The single rays are sent out from points on the opposite side of the computational grid, opposite to the detector pixels at which they are to be observed. The intensity of the ray at the start is either zero or set by the background radiation. When crossing a grid cell, the intensity of a ray is reduced corresponding to the optical depth $\tau$ of the dust in the cell by a factor $e^{-\tau}$ and increased by the thermal reemission radiation of the dust. This is done by solving the corresponding radiative transfer equation using the Runge-Kutta-Fehlberg method of order 4(5). When all cells along the path are processed, the total intensity of the ray is saved on the corresponding detector pixel. To ensure the applicability of the method for high optical depth, a recursive refinement for the integration step as well as the detector pixel size is applied (see,~\citealp{Ober2015}).

The time-dependent ray tracer is following this procedure, but saves the intensity of the rays for different time steps. This is done by stopping the calculation of the intensity along the ray after a path length $\delta l$ corresponding to the speed of light $c$ and the step width of a time step $\Delta t = \delta l/c$. The current total intensity of each ray is then scaled with the remaining optical depth towards the observer and projected onto the detector. For the next time step, the detector is replaced by a new detector on which the total intensity of the rays after the new time step is saved, and the intensity represented by the rays is again obtained by using the Runge-Kutta-Fehlberg method. This procedure is repeated for as many time steps as needed to account for the contributions of all volume elements along the path of each ray.

\begin{figure}[!ht]
    \centering
    \includegraphics[width=.75\linewidth]{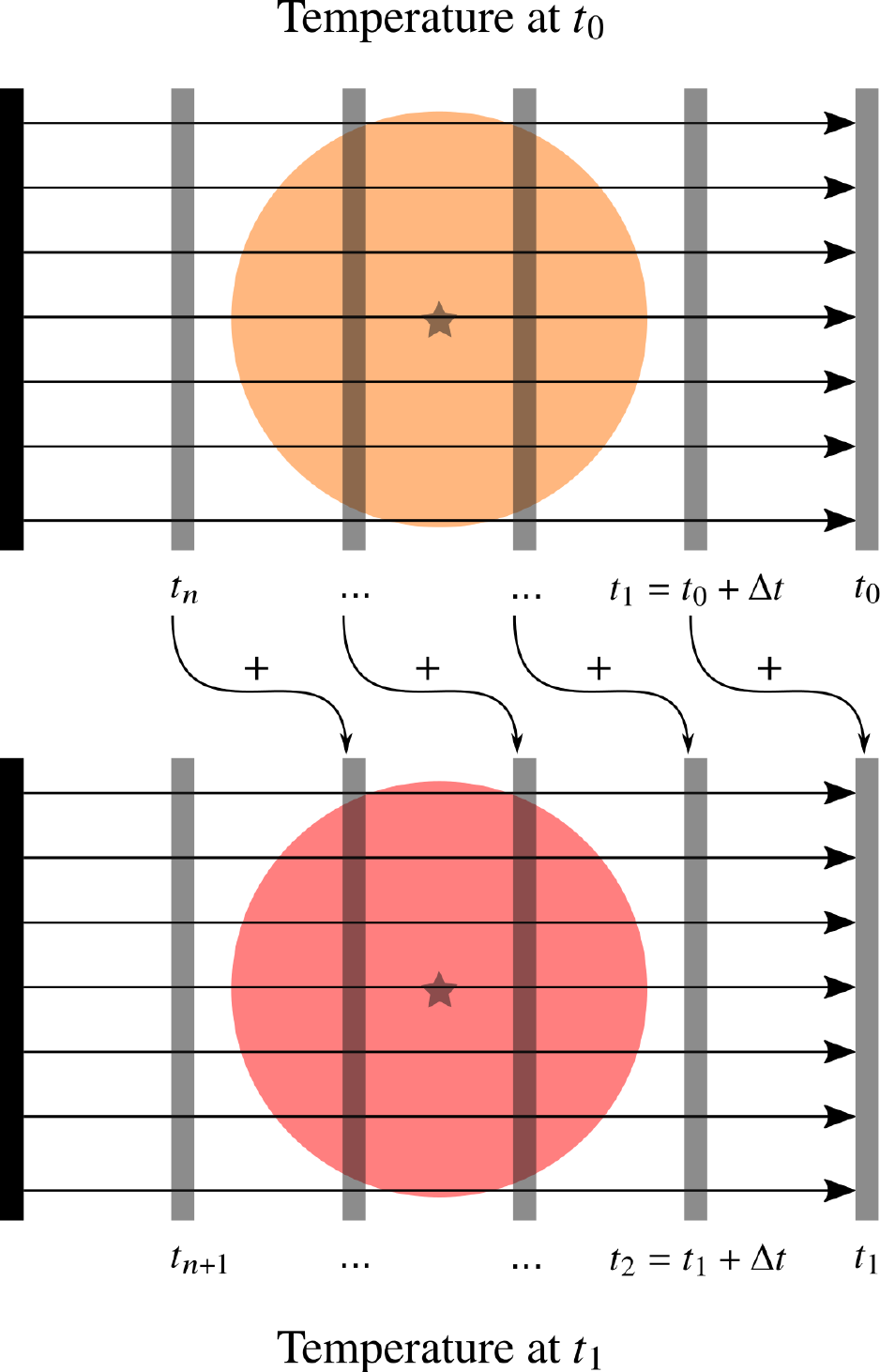}
    \caption{Sketch of the time-dependent ray tracing procedure for two different temperature distributions. The straight arrows denote the path of integration of each ray. The ray tracing simulation is performed based on the temperature distributions at time steps $t_0$ and $t_1$ separately. Detector images, that is, images at the position of the observer obtained for the same time step are added up.}
    \label{fig:raytracer}
\end{figure}

It should be noted that the temperature and thus the thermal reemission of each cell is changing with time. Therefore, one ray tracing simulation is only showing the contribution of the thermal reemission radiation of the temperature distribution of one time step. Consequently, every image calculated with the ray tracer only contains the contribution of the dust with the temperature distribution for that particular time step. Thus, some of the resulting images are showing the thermal reemission radiation at the same time steps, e.g., the first image simulated using the second temperature distribution shows the dust emission of the same time step as the second image of the simulation using the first temperature distribution. To get the final images, the images corresponding to the same time steps are added up. A sketch of the complete time-dependent ray tracing procedure for two different temperature distributions is shown in Fig.~\ref{fig:raytracer}.
A simple test of the corresponding light travel time effects for a spherical dust distribution can be found in Appendix~\ref{sec:sn_echo}.

\subsection{Time-dependent scattering}\label{subsec:timedep_sca}
As long as the dust density distribution of a model is not changed, e.g., due to dust sublimation, the distribution of the scattered stellar light does not depend on the temperature of the dust. Time-dependent scattering simulations can thus be performed independently of the temperature simulations. For this purpose, we modified the stationary Monte Carlo method for scattering simulations with peel-off first presented in~\cite{Yusef1984}.

Here, the basic handling of stellar emission and scattering are similar to the procedure described for the simulation of time-dependent temperature distributions in Sect.~\ref{subsec:timedep_temp}: For the wavelength of interest, a fixed number of photons packages are emitted from the source (e.g., star). The energy of the photon packages is assigned according to the spectrum of that source (e.g., Eq.~\ref{eq:alt_pp_en} for a stellar source). Next, the random directions of the photon packages and the optical depth of the first interaction are determined. The photon packages are then propagated through the model space until the optical depth of the first interaction is reached. in the case of a scattering event, a copy of the respective photon package (hereafter: peel-off photon) is sent to the detector. The energy of the peel-off photon is diminished in relation to the optical depth as well as the scattering probability to the observer. The scattered photon package is then propagated to the next optical depth of interaction. If a photon package leaves the model space, its energy is stored on a detector.

For the time-dependent scattering simulations, the path length of each photon package within the model space is integrated. In addition, the different distances to the detector due to different viewing
angles has to be considered. This is done by adding the remaining distance to the total path length
of a photon package after it leaves the model space. Peel-off photons inherit the path length of the
interacting photon package at the point of interaction and can thus be treated equally. The single
detector used for the stationary simulations is replaced by multiple detectors which cover different
intervals in time. The energy of a photon package leaving the model space will thus be saved on the detector that covers the time interval corresponding to the total path length of the photon package. Thus, using the speed of light $c$, we determine the index $n$ of a detector corresponding to the time step $t_n$ with step width $\Delta t$ for a photon package with total path length $l$ via: $l/c \in [t_n:t_n+\Delta t]$.

Changes of the properties of the emitting source are treated likewise by defining multiple sources which are emitting at different time steps. Since the time step currently assigned to a photon package is determined by the path length, the total path length of a photon package emitted by the $i$-th source at time $t_i$ has an offset of $l_0 = t_i\cdot c$. The procedure described above is then repeated until all photon packages of all sources left the model space.

\section{Tests}  \label{sec:test}

In the method described above, we introduced two parameters to handle time-dependent radiative transfer within the framework of an existing Monte Carlo approach, namely the temporal step width $\Delta$t and the number of photon packages per time step $N_{\textrm{ph}}$. In the following, we describe the influence of these parameters as well as the influence of the optical depth on the temperature distribution. This is done by simulating the heating process of a simple one-dimensional dust distribution around a centrally embedded illuminating and heating star for different values of each of the above parameters.

The dust mixture used for all models corresponds to the typical composition of the ISM with 62.5\% astronomical silicate and 37.5\% graphite, for which we apply the $\frac{1}{3}$-$\frac{2}{3}$ relation for parallel and perpendicular orientations (see,~\citealp{DraineMalhotra1993}). The grain size distribution follows \cite{MathisRumplNordsieck1977} and can thus be described by a power law $n(s) \propto s^{-3.5}$ for grain sizes between $5\;$nm and $250\;$nm. The optical properties of the dust are derived using the wavelength-dependent refractive indices by \cite{Draine1984} and \cite{Laor1993}. The calorimetric data is obtained by applying the models for silicate and graphite presented in~\cite{DraineLi2001}.
To save computational time, we assume mean values for all dust grain parameters. Following \cite{Wolf2003b} they are weighted with respect to the abundances of the chemical components and the grain number density as well as the grain size distribution. To justify this approach, we compared temperature simulations using multiple and single grain sizes to the respective simulations using the stationary method. We found that the temperature distributions calculated with weighted mean values are deviating less than 1\% from temperature distributions simulated with unaveraged values.

\subsection{Temporal step width $\Delta$t}\label{subsec:dt}
\begin{figure}[!ht]
    \centering
    \includegraphics[width=1.\linewidth]{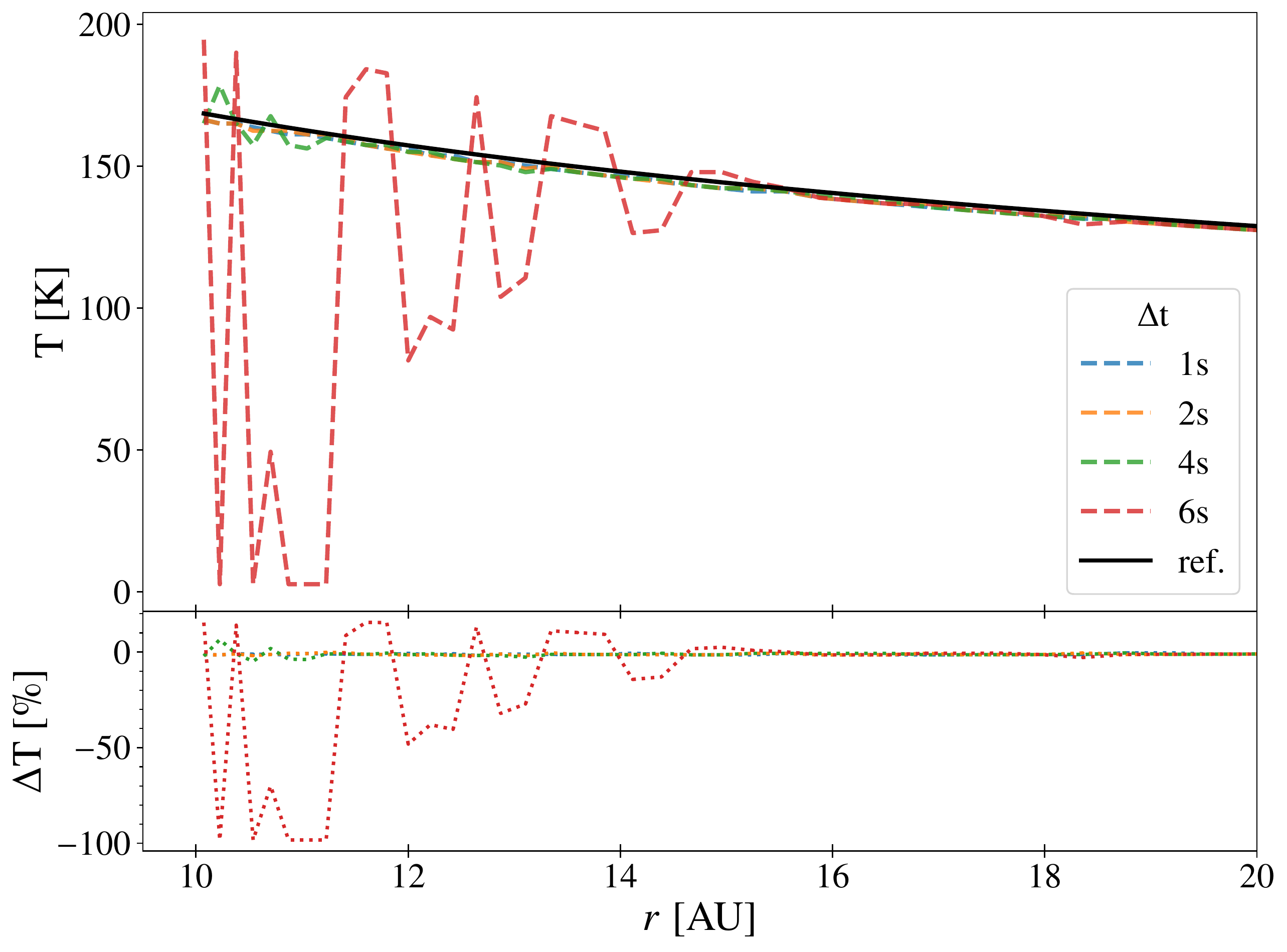}
    \caption{Resulting temperature distribution (top) and relative difference (bottom) of the 1D model discussed in Sect.~\ref{subsec:dt} with different temporal step widths $\Delta t$ for the time-dependent (dashed) and stationary (solid) method. The number of photon packages per step is set to $10^3$, the total number of photon packages used for the stationary simulation is $10^8$, respectively. Lines for radii larger than 20\,AU are overlapping and are therefore not shown.}
    \label{fig:1D_dt}
\end{figure}

In Sect.~\ref{subsec:phys} we have shown that the temporal step width is a crucial parameter for the goodness of the estimator of the enthalpy and thus for the simulation of the temperature distribution. To illustrate the effect of different values of this parameter, we simulated the temperature distribution of a simple one-dimensional dust distribution with constant density (inner radius $R_{\rm in} = 10$\,AU, outer radius $R_{\rm out} = 100$\,AU, optical depth $\tau_{\rm V}=0.1$) and an embedded TTauri-like star (hereafter: 1D model) and compared it to the resulting temperature distribution of the stationary method of \texttt{POLARIS}. Since the stellar parameters had been fixed during the simulation, the dust in the grid heated up until the equilibrium temperature was reached. The simulation was started with a dust temperature of 2.7\,K for all cells and stopped when the temperature of the dust in each cell is no longer changing within a $\Delta T$ given by the bin size of the temperature grid (logarithmically distributed between 2.7\,K and 2500\,K). To ensure that the time interval is long enough to reach this equilibrium, the simulations covered at least twice the light traveling time through the grid. Thus, a total simulation time of $10^5$\,s was chosen for the given model. To rule out the influence of the number of photon packages per time step on the simulations, we fixed $N_{\rm ph}$ at $10^3$ for all simulations. Thus, the total number of photon packages emitted over the simulation time for the smallest temporal step width was $10^8$. Therefore, this number of photon packages was used for the simulation of the stationary reference temperature distribution. An overview of all model parameters can be found in Tab.~\ref{tab:simple}. The resulting temperature distributions of the simulations with step widths $\Delta t$ = 1\,s, 2\,s, 4\,s and 6\,s are shown in Fig.~\ref{fig:1D_dt}.

\begin{table}[!h]
  \begin{center}
    \caption{Overview of the parameter of the spherical 1D model discussed in Sect.~\ref{sec:test}.}
    \label{tab:simple}
    \begin{tabular}{llc}
        \hline
        \hline
        \rule{0pt}{2ex}
        \textbf{Parameter} & & \textbf{Value}\\
        \hline
        \rule{0pt}{3ex}
        Inner radius & $R_{\rm{in}}[$AU$]$ & 10 \\
        \rule{0pt}{1ex}
        Outer radius & $R_{\rm{out}}[$AU$]$ & 100 \\
        \rule{0pt}{1ex}
        Total dust mass & $M_\text{dust}$[M$_\odot$] & $1.7 \cdot 10^{-6}$\\
        \rule{0pt}{1ex}
        Optical depth & $\tau_{\rm V}$ & 0.1\\
        \rule{0pt}{1ex}
        Source radius & $R_\star$[R$_\odot$] & 2\\
        \rule{0pt}{1ex}
        Source temperature & $T_{\star}$[K]  &  4000\\
        \rule{0pt}{1ex}
        Number of cells in & &\\
        \rule{0pt}{1ex}
        ... radial direction & $N_{r}$ &  $100$\\
        \rule{0pt}{1ex}
        ... azimuthal direction & $N_{\phi}$ & $1$\\
        \rule{0pt}{1ex}
        ... polar direction & $N_{\theta}$ &  $1$\\
        \hline
    \end{tabular}
  \end{center}
\end{table}
While the temperature distributions obtained for the temporal step widths $\Delta t = 1$\,s and 2\,s converged to the stationary temperature distribution, the temperature distributions calculated with $\Delta t = 4$\,s and 6\,s are oscillating in the inner region. This can be explained by calculating the cooling time $t_{\rm c}$ for the cells with the highest dust temperature using Eq.~\ref{eq:cool}. If the temporal step width is close to the cooling time, these cells will cool almost to the ground temperature of 2.7\,K in a single time step. Thus, the estimated emission rate of the dust will be low compared to the absorption rate. Since absorption and emission rates are considered constant within one time step, this leads to an overestimation of the dust enthalpy given by Eq.~\ref{eq:nextu} and thus an overestimated dust temperature. These cells with higher dust temperatures will cool even faster. The dust temperatures of the affected cells are then oscillating between very high and low values. Since the dust is emitting photon packages in every time step, and overestimated dust temperatures lead to an overestimated energy of the emitted photon packages, the oscillation propagates through the grid. The larger the temporal step width compared to the cooling time, the stronger the oscillation. For this reason, the step width should always be chosen according to the cooling time of the estimated maximum dust temperature of the simulation.

\subsection{Number of photons per time step $N_{\textrm{ph}}$}\label{subsec:nph}
The number of photon packages is a crucial parameter for every numerical approach to the radiative transfer problem that makes use of the Monte Carlo method. in the case of time-dependent simulations, this has to be taken into account for each individual time step. The effect of an increase of the number of photon packages on the computational time of each simulation is stronger than in the stationary case because all photon packages need to be stored and loaded for every time step. Furthermore, the total number of photon packages processed in every time step will increase until the number of photon packages leaving the grid equals the number of photon packages that are emitted. The computation time of a single time step is thus increasing linearly. It is therefore crucial to evaluate the effect of the number of photon packages per time step on the resulting temperature distribution.

\begin{figure}[!ht]
    \centering
    \includegraphics[width=1.\linewidth]{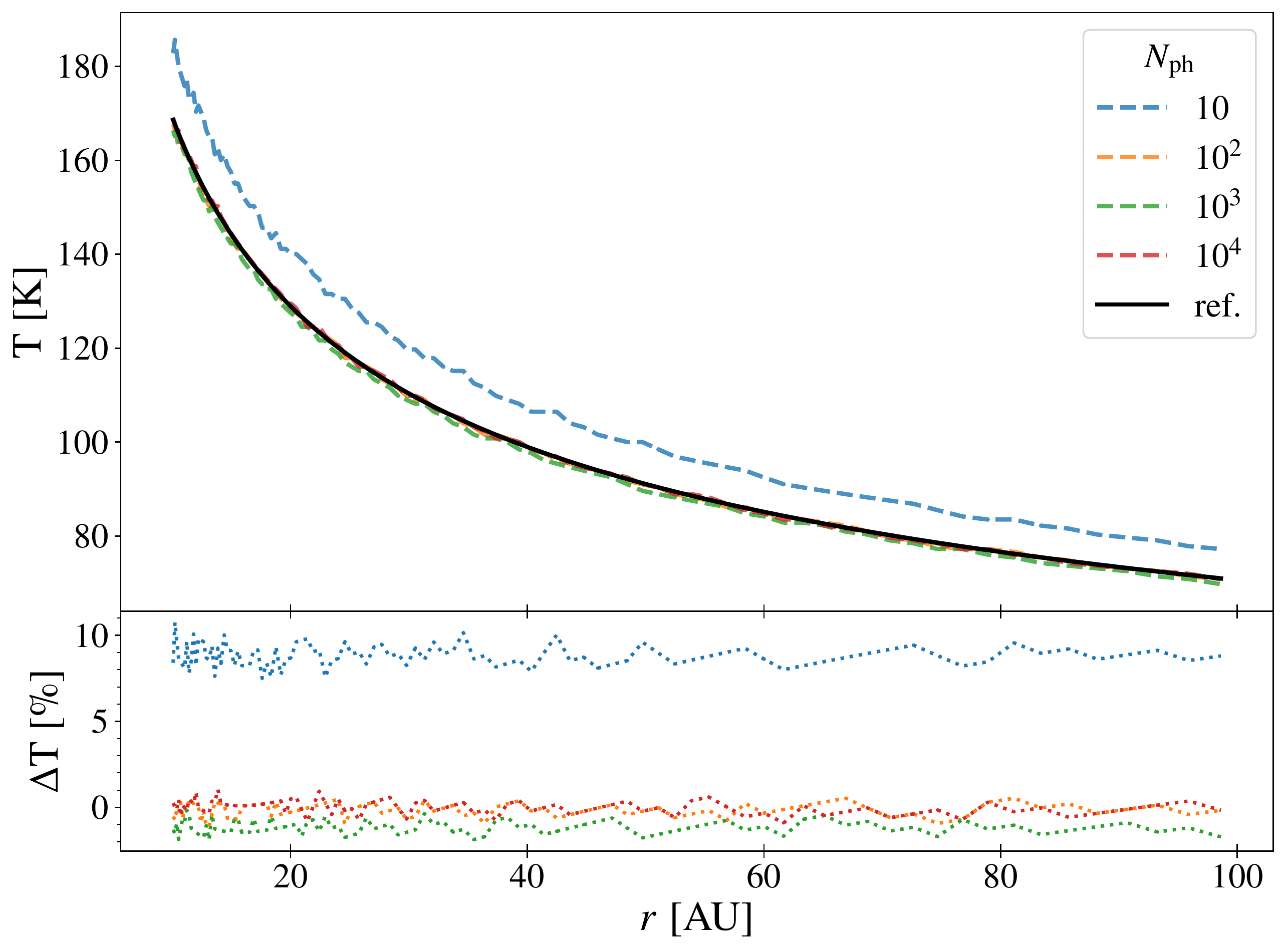}
    \caption{Resulting temperature distribution (top) and relative difference (bottom) of the 1D model discussed in Sect.~\ref{subsec:nph} with different number of photon packages per time step $N_{\rm ph}$ for the time-dependent (dashed) and stationary (solid) method. The temporal step width is set to $2$\,s, the total number of photon packages used for the stationary simulation is $10^8$, respectively.}
    \label{fig:1D_nrph}
\end{figure}

For this purpose, we performed a series of simulations for the simple 1D model presented in Sect.~\ref{subsec:dt} with a fixed temporal step width of $\Delta t = 2$\,s and different number of photon packages per time step $N_{\rm ph}=10, 10^2, 10^3, 10^4$. The temperature distribution obtained with the stationary approach using $10^8$ photon packages in Sect.~\ref{subsec:dt} was used as reference. The resulting temperature distributions and relative differences are shown in Fig.~\ref{fig:1D_nrph}.
It can easily be seen, that the temperature distribution gets smoother for higher numbers of photon packages. Furthermore, a value of $N_{\rm ph} = 10$ leads to an overestimation of the dust temperature in the inner cells.

Since the probability of an absorption event in the case of $\tau_{\rm V}=0.1$ is sufficiently small, it can be assumed that most of the photon packages will not be absorbed, that is their energy will be carried through all cells until the photon package leaves the model space. However, some photon packages will undergo an absorption event, that is their further transfer through the grid will be stopped. Following the approach to handle absorption discussed in Sect.~\ref{subsec:timedep_temp} a photon package is only contributing energy to the enthalpy of the dust by continuous absorption along its path through the grid. Thus, after an absorption event, the radiation field in the following cells is weakened. In the case of a very low number of photon packages per time step ($N_{\rm ph} < N_{\rm Cells}$) nearly no photon package will be absorbed and the whole energy emitted by the source in every time step contributes to the dust enthalpy of every cell. Thus, the dust temperature of each cell is overestimated. However, in the case of a single absorption event, the impact on the local radiation field will be larger (e.g., in the case of $N_{\rm Ph} = 10$ roughly 10\% of the energy distributed in one time step is taken out), leading to large variations of the temperature distribution. In the case of higher optical depth, that is if absorption is more probable, the dust temperature will be underestimated. The same applies for the wavelength selection, which is also done by using a probability density function (see, Eq.~\ref{eq:wave_prob}). The flux at wavelength where the emission probability is highest will be overestimated, while the flux at wavelength where the emission probability is lowest will be underestimated.

\subsection{Optical depth}\label{subsec:1D_depth}
\begin{figure}[!ht]
    \centering
    \includegraphics[width=1.\linewidth]{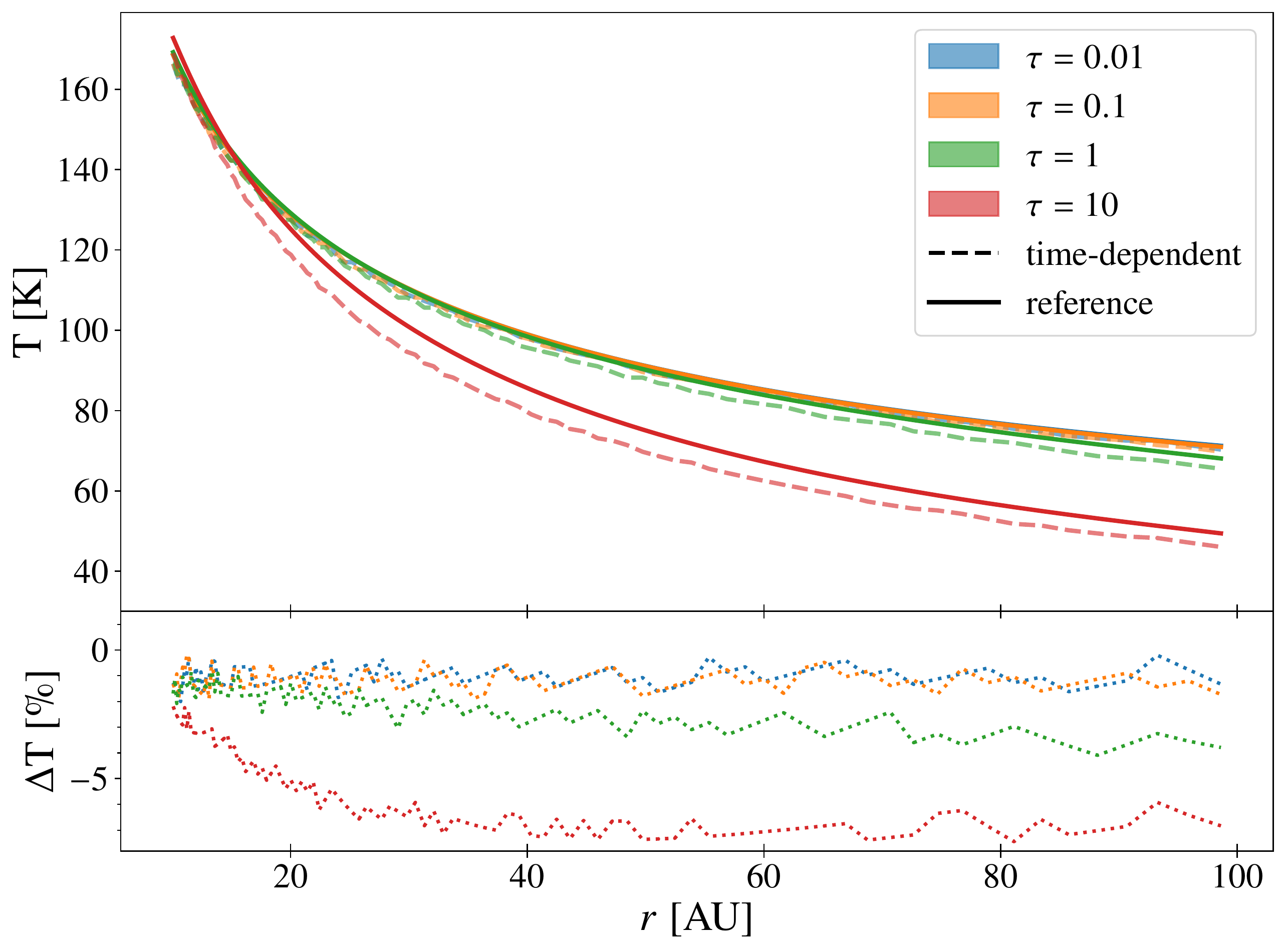}
    \caption{Resulting temperature distributions (top) and relative differences (bottom) of the 1D model for different optical depths $\tau_{\rm V}$ discussed in Sect.~\ref{subsec:1D_depth} for the time-dependent (dashed) and stationary (solid) method. in the case of time-dependency, the temporal step width is fixed at 2\,s and the number of photon packages per time step is $10^3$. The total number of photon packages used for the stationary simulations is $10^8$.}
    \label{fig:1D_midtemp}
\end{figure}

To examine the behavior of the optical depth on the time-dependent temperature simulation, we performed simulations of models with different optical depths ($\tau_{\rm V}=0.01, 0.1, 1, 10$ as seen from the central star). We fixed the temporal step width and the number of photons per time step to values that lead to well converged temperature distributions in Sect.~\ref{subsec:dt} and \ref{subsec:nph}, that is, $\Delta t = 2$\,s and $N_{\textrm{ph}} = 10^3$. For every optical depth, a reference temperature distribution was simulated using the stationary method for temperature distributions. The resulting temperature distributions of the time-dependent simulations and the relative differences to the respective reference temperature distributions after a total simulation time of $10^5$\,s are shown in Fig.~\ref{fig:1D_midtemp}.
It can be seen that the temperature distributions of the models up to $\tau_{\rm V}=1$ are in very good agreement with the temperature distributions of the stationary simulations. in the case of the $\tau_{\rm V}=10$ model, the dust temperature is always lower than the reference temperature distribution. This is caused by a slight discrepancy between the optical and calorimetric data used in the time-dependent temperature simulations, as well as some numerical constraints that are further discussed in Sect.~\ref{sec:disc}. However, the temperature distribution of the time-dependent simulation is deviating less than 8\% from the respective temperature distribution of the stationary approach and is therefore considered consistent.

\section{Examples}  \label{sec:exam}
Since the goal of our method is the simulation of young stellar objects with circumstellar disks, we now consider the radiative transfer of a model of a very low-mass circumstellar disk. The model description as well as a comparison between time-dependent and stationary simulation of the temperature distribution is given in Sect.~\ref{subsec:2D_depth}. Time-dependent simulations of the scattered light as well as the thermal reemission radiation for a simple luminosity outburst of the stellar source are presented in Sect.~\ref{subsec:2D_burst}.

\subsection{Circumstellar disk model}\label{subsec:2D_depth}
The density distribution of the disk is following the considerations of~\cite{ShakuraSunjaev} and has been successfully used for modeling circumstellar disks in various studies (e.g.,~\citealp{Pinte2008};~\citealp{Ratzka2009};~\citealp{Brunngraeber2016}). It is given by:
\begin{equation}
    \label{eq:dens_dist}
    \rho (r,z) = \rho_0 \left(\frac{r}{R_0}\right)^{-\alpha} \exp\left[ - \frac{1}{2} \left(\frac{z}{h(r)}\right)^2\right],
\end{equation}
with the cylindrical coordinates $(r,z)$, the density scaling parameter $\rho_0$ determined by the total disk mass, the radial density profile parameter $\alpha$ and $h(r)$ the scale height, which can be written as
\begin{equation}
    \label{eq:scale_height}
    h(r) = h_{\rm ref} \left(\frac{r}{R_0}\right)^\beta.
\end{equation}
Here, $R_0$ is the radius of the reference scale height $h_{\rm ref}$. The parameter $\beta$ determines the disk flaring. We applied this density distribution for a model with an inner radius of $R_{\rm{in}} = 20$\,AU, outer radius $R_{\rm{out}} = 100$\,AU and moderate flaring ($\beta = 1.125$ and $h_{\rm{ref}} = 10$\,AU).
\begin{figure*}[!ht]
    \centering
    \includegraphics[width=1.\linewidth]{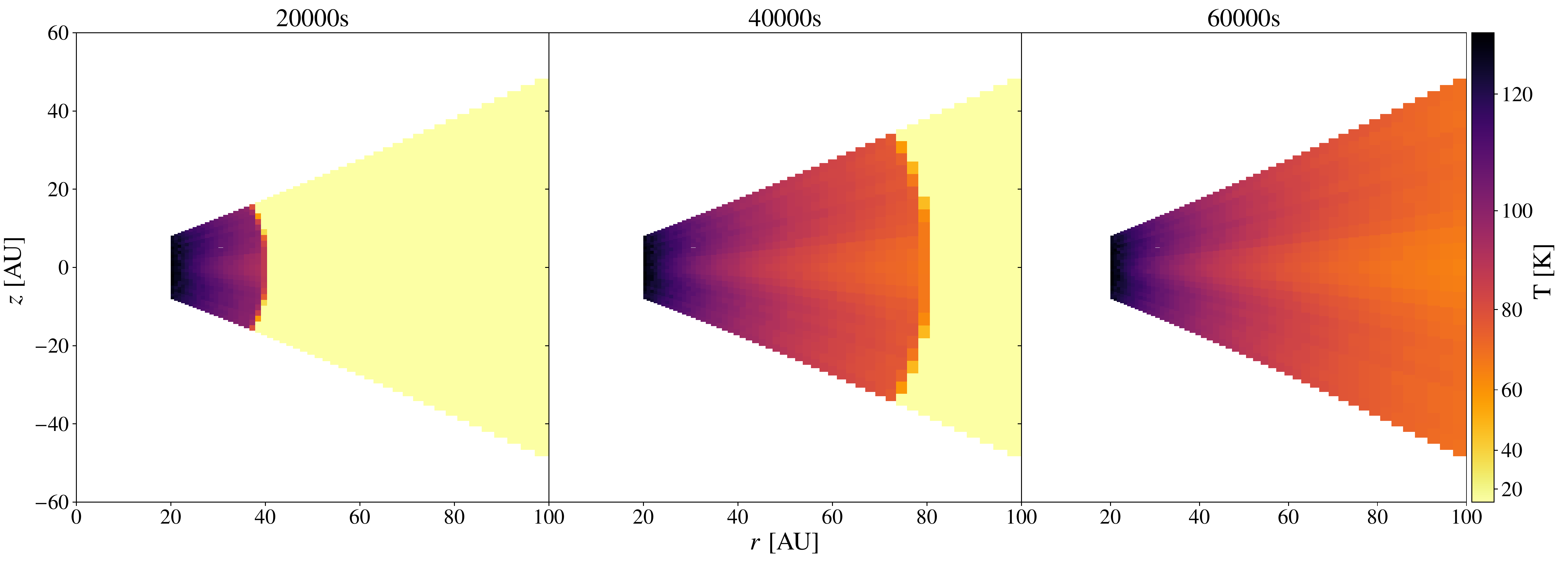}
    \caption{Vertical cuts through the temperature distribution of the time-dependent disk model after 20\,000, 40\,000 and 60\,000\,s described in Sect.~\ref{subsec:2D_depth} with $\tau=1$ and temporal step width 8\,s and $10^5$ photons per time step.}
    \label{fig:2D_heat}
\end{figure*}
\begin{table}[!h]
  \begin{center}
    \caption{Overview of the parameters of the disk model discussed in Sect.~\ref{subsec:2D_depth}.}
    \label{tab:disk}
    \begin{tabular}{llc}
        \hline
        \hline
        \rule{0pt}{2ex}
        \textbf{Parameter} & & \textbf{Value}\\
        \hline
        \rule{0pt}{3ex}
        Inner radius & $R_{\rm{in}}[\rm{AU}]$ & 20 \\
        \rule{0pt}{1ex}
        Outer radius & $R_{\rm{out}}[\rm{AU}]$ & 100 \\
        \rule{0pt}{1ex}
        Source radius & $R_\star$[R$_\odot$] & 2\\
        \rule{0pt}{1ex}
        Source temperature & $T_{\star}$[K]  &  4000\\
        \rule{0pt}{1ex}
        Radial density exponent & $\alpha$  &  2.625\\
        \rule{0pt}{1ex}
        Disk flaring & $\beta$  &  1.125\\
        \rule{0pt}{1ex}
        Reference scale height & $h_{\rm{ref}}[\rm{AU}]$  &  10\\
        \rule{0pt}{1ex}
        Temporal step width & $\Delta t$[s]  &  8\\
        \rule{0pt}{1ex}
        Number of photons per step & $N_{\rm{ph}}$ &  $10^5$\\
        \rule{0pt}{1ex}
        Number of cells in & &\\
        \rule{0pt}{1ex}
        ... radial direction & $N_{\rm{r}}$ &  $50$\\
        \rule{0pt}{1ex}
        ... axial direction & $N_{\rm{z}}$ &  $23$\\
        \rule{0pt}{1ex}
        ... azimuthal direction & $N_{\phi}$ &  $1$\\
        \hline
    \end{tabular}
  \end{center}
\end{table}

To make sure that the two-dimensional density structure has no effect on the time-dependent temperature distribution, we performed simulations of the disk model with midplane optical depths of $\tau_{\rm V}=0.01, 0.1, 1$, and 10. We then compared these to the temperature distributions simulated with the stationary method. The temporal step width ($\Delta t = 8$\,s), number of photons per time step ($N_{\rm ph} = 10^5$) and the total simulation time ($10^5$\,s) were chosen according to the criteria discussed in Sect.~\ref{sec:test}. Corresponding to the total number of photon packages used for the time-dependent temperature simulations, the number of photon packages applied for the stationary simulation of each model was $10^9$. An overview of all parameters of the disk model can be found in Tab.~\ref{tab:disk}.

\begin{figure}[!ht]
    \centering
    \includegraphics[width=1.\linewidth]{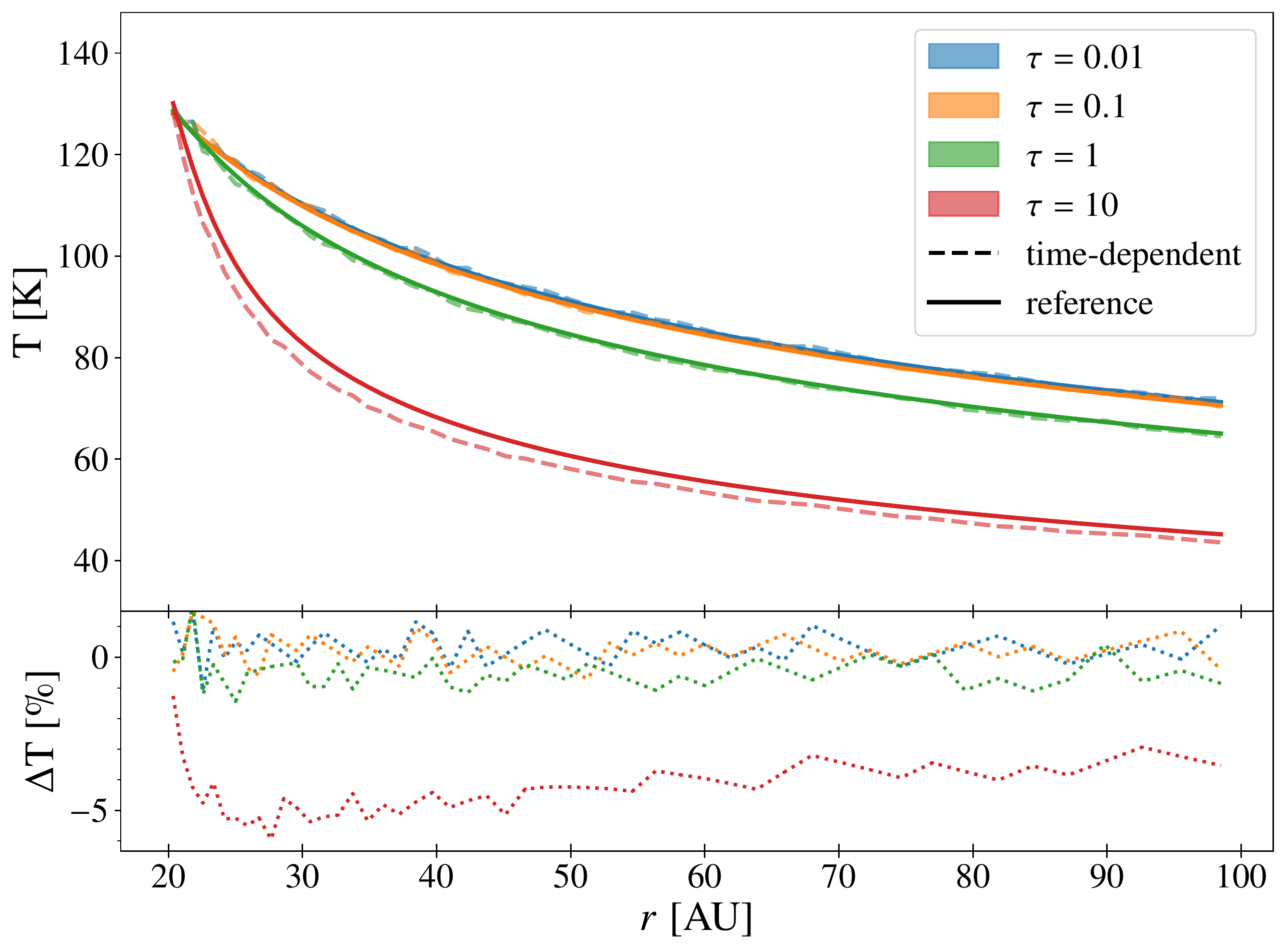}
    \caption{Resulting midplane temperature distributions (top) and relative differences (bottom) of the disk model with different optical depth $\tau_{\rm V}$ discussed in Sect.~\ref{subsec:2D_depth} for the time-dependent (dashed) and stationary (solid) method. in the case of time-dependency, the temporal step width is fixed at 8\,s and the number of photon packages per time step is $10^5$. The total number of photon packages used for the stationary simulations is $10^8$.}
    \label{fig:2D_midtemp}
\end{figure}

\begin{figure}[!ht]
    \centering
    \includegraphics[width=1.\linewidth]{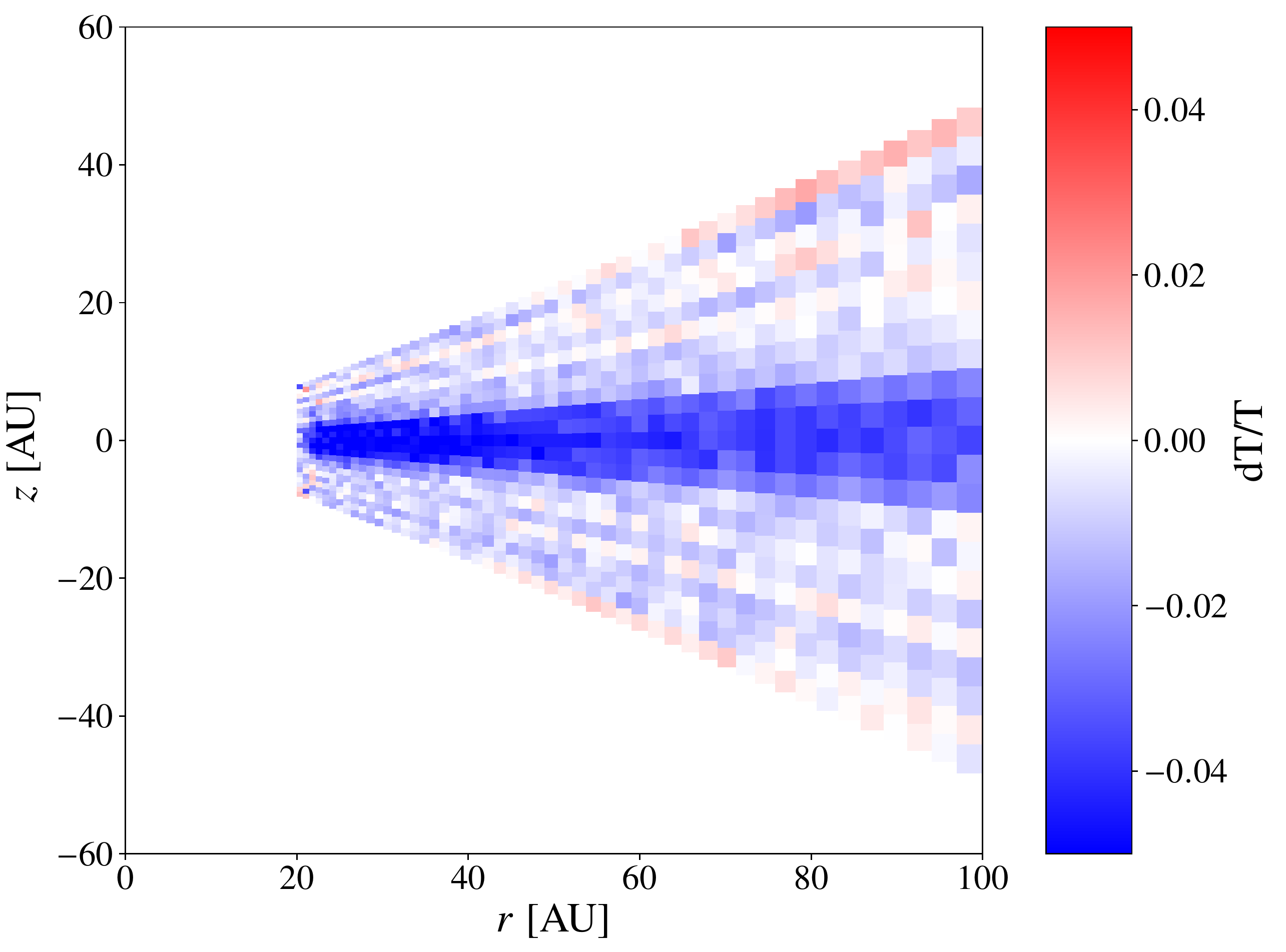}
    \caption{Difference plot of a vertical cut through the dust temperature distribution of the disk model with $\tau=10$ described in Sect.~\ref{subsec:2D_depth}. The reference temperature distribution is calculated with the stationary method using $10^8$ photon packages.}
    \label{fig:2D_diff}
\end{figure}

A selection of vertical cuts through the temperature distribution of the time-dependent simulation for an optical depth $\tau_{\rm V}=1$ for different time steps, that is different stages of the heating process, is depicted in Fig.~\ref{fig:2D_heat}. The resulting temperature distributions and relative differences of the midplane temperature for all models are shown in Fig.~\ref{fig:2D_midtemp}.
The trend of the temperature distribution with different optical depth of the 1D model discussed in Sect.~\ref{subsec:1D_depth} can also be found for the disk model. The time-dependent temperature distributions with optical depths of $\tau_{\rm V}=0.01, 0.1, 1$ are in good agreement with the temperature distributions of the stationary simulations. Since the optical depth is highest in the midplane of the disk model, we can assume that the temperatures of the entire disk are equally well fitting.
In the case of $\tau_{\rm V}=10$, the temperature distribution in the midplane is deviating by about 5\% from that in the midplane of the stationary simulation. To check whether this also affects the dust temperature of cells above and below the midplane, a plot of the difference of the dust temperature distribution between stationary and time-dependent simulation for all grid cells is given in Fig.~\ref{fig:2D_diff}.
It can be seen that the deviation of the dust temperature of the cells above and below the midplane is always below 2\% and thus comparable to the level of the statistical noise of the simulation.

\begin{figure}[t]
    \centering
    \includegraphics[width=1.\linewidth]{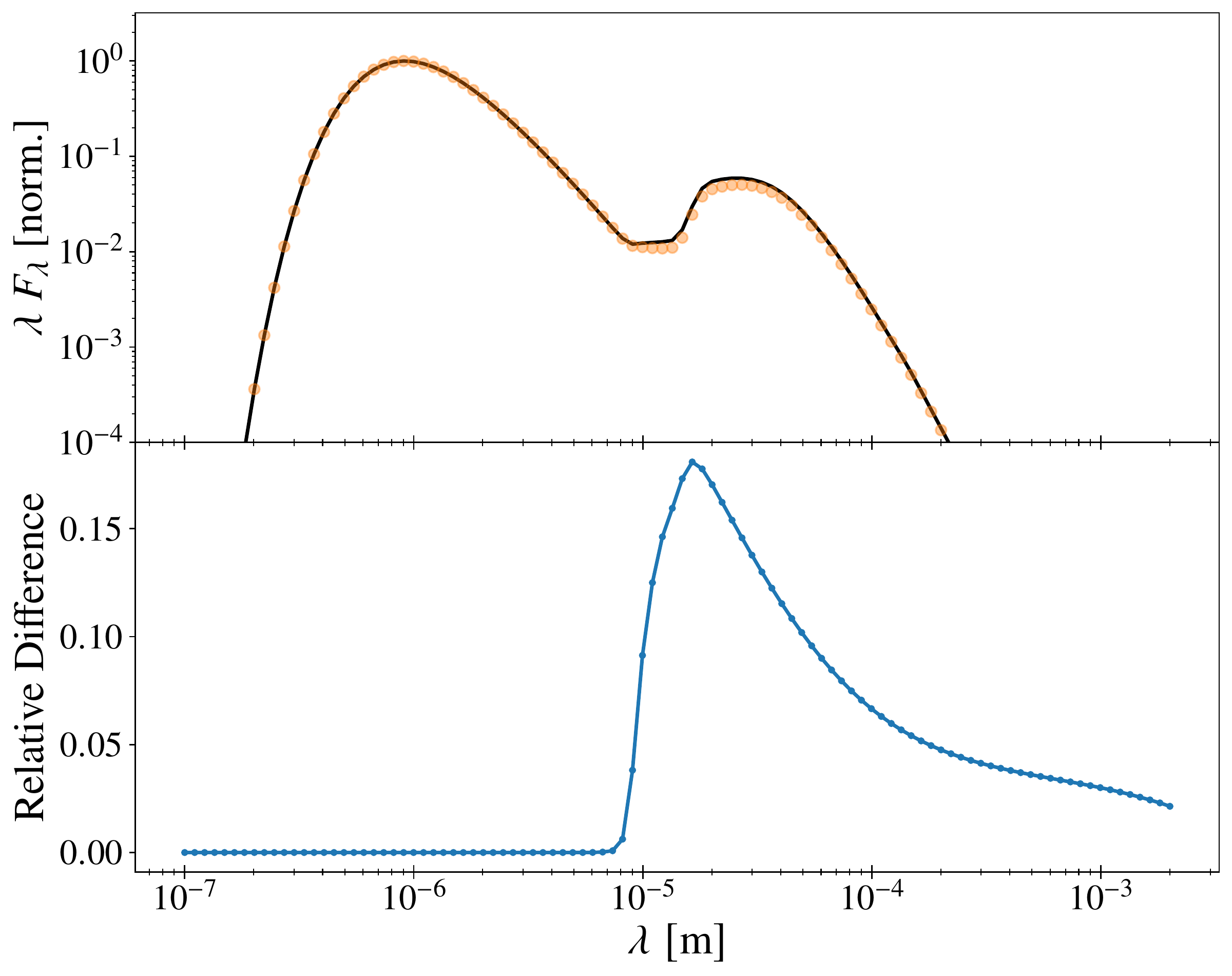}
    \caption{Normalized SED (top) and relative difference (bottom) of the time-dependent temperature distribution of the disk model with $\tau=10$ discussed in Sect.~\ref{subsec:2D_depth} (dotted). The reference SED (solid) is calculated using the temperature distribution of the stationary method.}
    \label{fig:2D_sed}
\end{figure}

To make sure that the corresponding spectral energy distribution (SED) is not significantly affected by the underestimated dust temperature in the midplane, we performed simulations of the scattered light and thermal reemission radiation for the time-dependent temperature distribution as well as for the reference temperature distribution. The resulting SEDs as well as the relative difference between the SEDs calculated using the time-dependent and the stationary temperature distribution are shown in Fig.~\ref{fig:2D_sed}.

While the overall shape of the SED is well reproduced, the difference plot reveals a peak difference of about 15\% for wavelengths of around 20\;$\upmu$m. This is close to the shortest wavelength at which the thermal reemission radiation of the dust for the given temperature distribution becomes significant. The difference plots in Fig.~\ref{fig:2D_midtemp} and \ref{fig:2D_diff} show that the dust temperature of the cells in the inner few AU of the midplane is underestimated. These cells correspond to the cells with the highest dust temperature determined with the stationary method. The spectrum of the thermal reemission radiation of the dust with the time-dependent temperature distribution is thus shifted to longer wavelengths compared to the spectrum calculated with the stationary method. However, this only affects the radiation of wavelengths corresponding to the thermal reemission of the few cells with the highest dust temperature. Therefore, the flux of most of the wavelengths larger than 10\;$\upmu$m is showing not more than 10\% difference to the reference SED.

\subsection{2D disk outburst}\label{subsec:2D_burst}
Lastly, we performed simulations of a model for an outburst of a star with a circumstellar disk. For this, we used the disk model described in Sect.~\ref{subsec:2D_depth} (see also Tab.~\ref{tab:disk}). For the sake of simplicity and to save computational time, we chose a low optical depth of $\tau_{\rm V} = 0.01$. In this case the dust temperature of the cells are converged after about 50\,000\,s, that is roughly the time light needs to travel once through the model space. The luminosity of the central star was then increased by a factor of 4 for a duration of 10\,000\,s. The total simulation time was set to 120\,000\,s to make sure, that all grid cells return to their previous dust temperature after the outburst.

\begin{figure*}[!h]
    \centering
    \includegraphics[width=1.\linewidth]{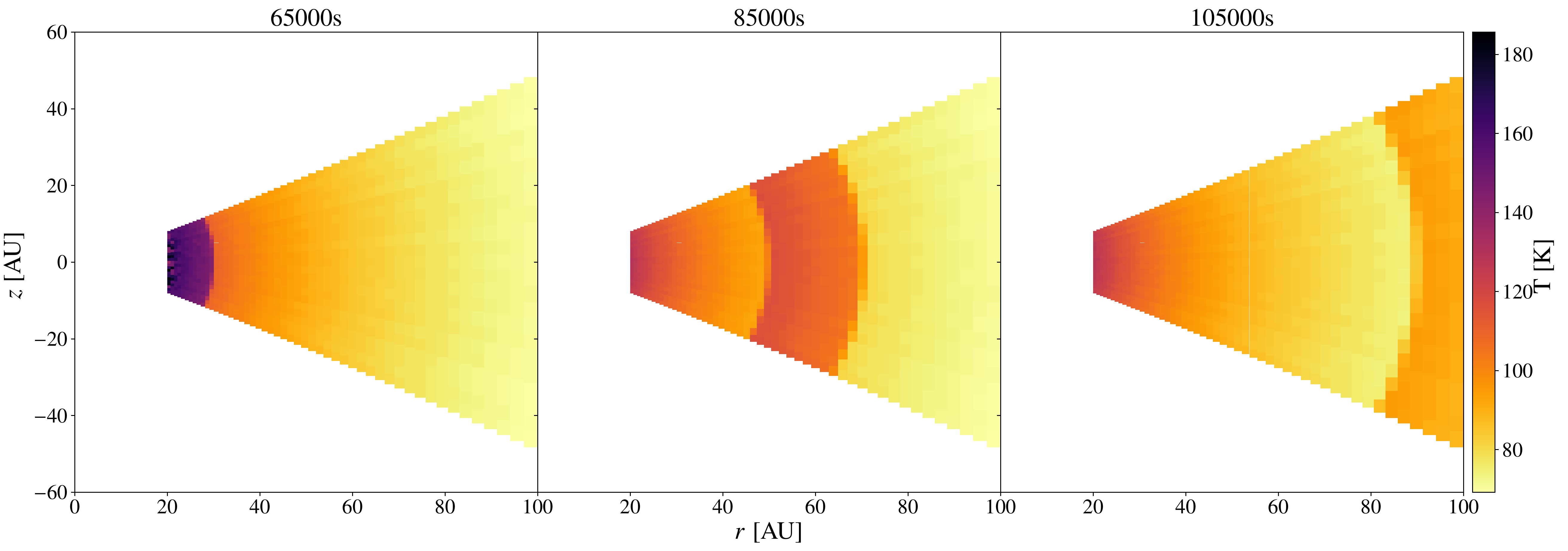}
    \caption{Vertical cut through the temperature distribution of the outburst disk model after 65\,000, 85\,000 and 105\,000\,s described in Sect.~\ref{subsec:2D_burst}. The temporal step width of the simulation is set to 4\,s; $10^5$ photon package are emitted per time step.}
    \label{fig:2D_burst}
\end{figure*}

The temperature simulation was performed using a temporal step width of 4\,s and $10^5$ photon packages per time step. The temporal step width was chosen such that the cooling time of the expected highest dust temperature is not reached within one step. Three vertical cuts through the temperature distribution of the disk model for different time steps during the outburst are shown in Fig.~\ref{fig:2D_burst}.
It can easily be seen that a heat wave corresponding to the outburst in the source luminosity is traveling through the grid in radial direction. Furthermore, the thermal response times of the dust are comparably small, that is the dust of a cell is heating or cooling to the equilibrium temperature within one time step, leading to a discrete form of the heat wave.

Furthermore, we investigated the scattered stellar light and thermal reemission radiation of the dust as observables of this system. In particular, we simulated the thermal reemission radiation of the N band (10\;$\upmu$m) and the scattered stellar light of the V band (550\;nm) using the respective time-dependent algorithms for ray tracing and scattering described in Sect.~\ref{subsec:timedep_ray} and Sect.~\ref{subsec:timedep_sca}. In the case that the observed disk is inclined, it is necessary to take the different light traveling times to the observer into account. To illustrate this effect, the intensity maps were derived for an inclination of $25^\circ$ as well as for the face-on case ($0^\circ$). Since the scattering and reemission simulations are not sensitive to the temporal step width, the step width was set to 1000\,s and a total number of photon packages per time step of $10^8$ were used for scattering and reemission, respectively. In order to avoid an impact of the initial heating process, all simulations were started once a steady state was reached. To rule out an offset of the light curves of the scattered and reemitted radiation caused by additional light travel times due to different positioning of the detector in the scattering and ray tracing treatments, the start of the outburst was set to 60\,000\,s in both cases.

\begin{figure}[!ht]
    \centering
    \includegraphics[width=1.\linewidth]{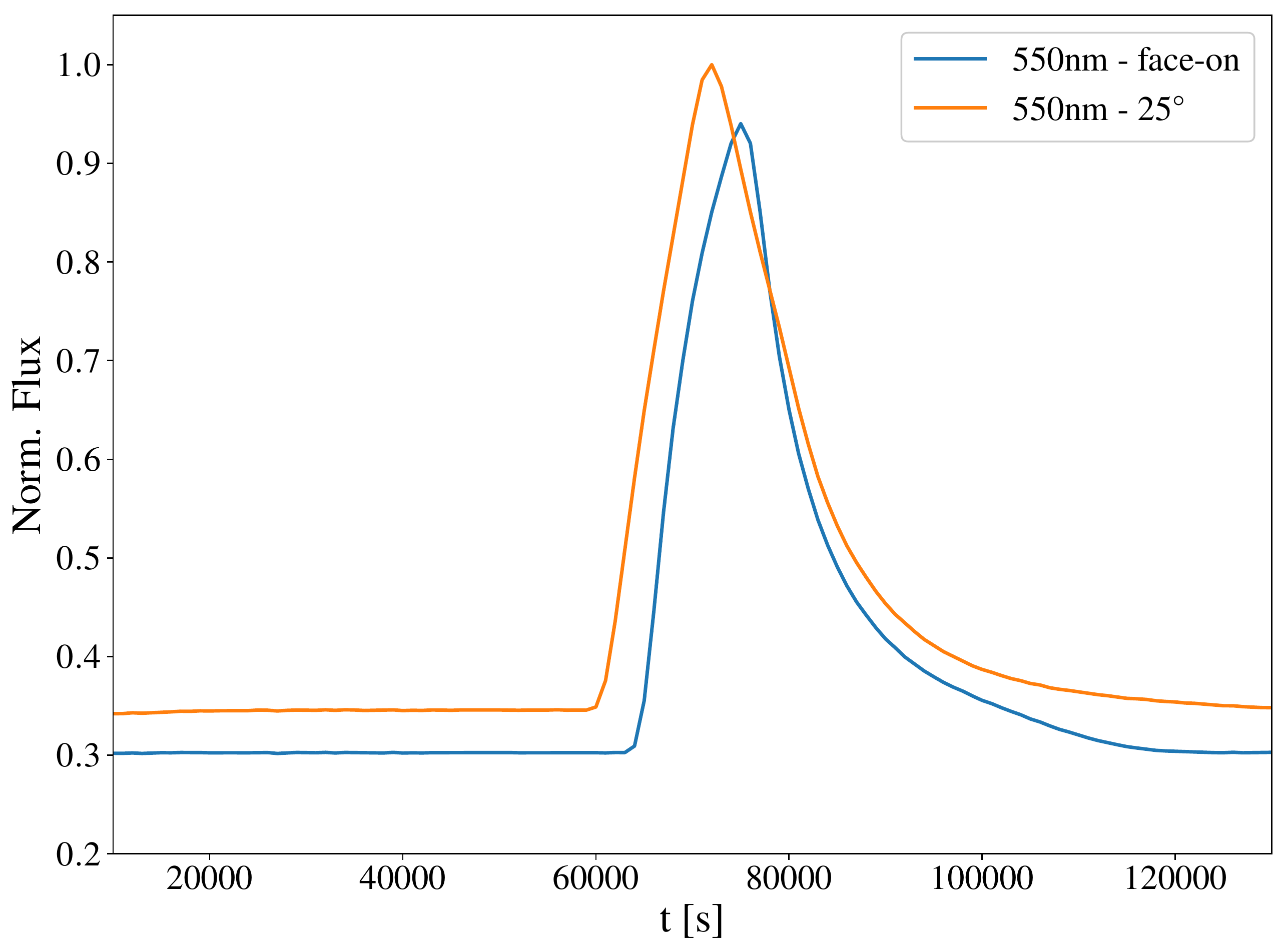}
    \caption{Light curve of the scattered stellar light (550\;nm) of the outburst model discussed in Sect.~\ref{subsec:2D_burst}. The duration of the outburst is 10\,000\,s; the total simulation time is 120\,000\,s. During the outburst, the stellar luminosity is increased by a factor of 4. The flux is normalized with respect to the maximum of the integrated flux of all images of the different time steps.}
    \label{fig:burst_sca}
\end{figure}

\begin{figure*}[!ht]
    \centering
    \includegraphics[width=1.\linewidth]{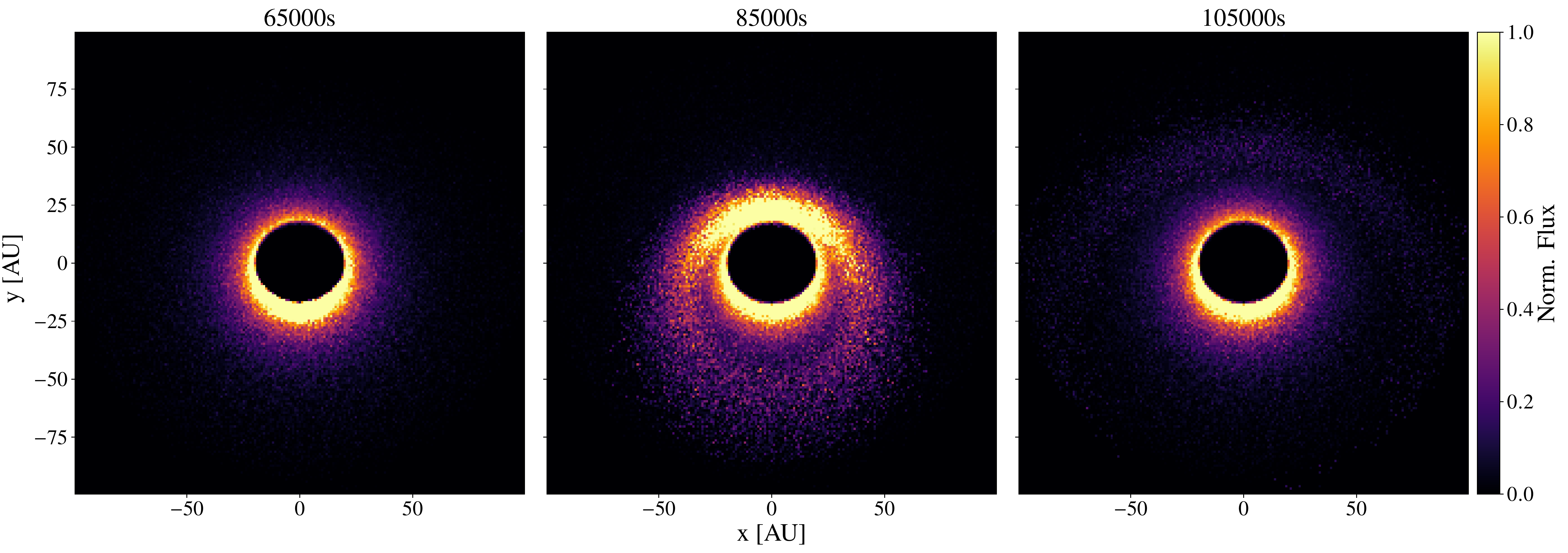}
    \caption{Images of the scattered light of the outburst model after 65\,000, 85\,000 and 105\,000\,s for 550\;nm described in Sect.~\ref{subsec:2D_burst} (temporal step width: 1000\,s; $10^8$ photon packages per time step). The flux is normalized with respect to the maximum flux of all images of the different time steps.}
    \label{fig:2D_burst_sca}
\end{figure*}

The resulting light curve for the scattered light is shown in Fig.~\ref{fig:burst_sca}. A sample of the corresponding scattering images for the inclined disk can be found in Fig.~\ref{fig:2D_burst_sca}.
There are three effects that can be seen in the light curves: First, the overall flux is higher in the case of the inclined disk. This is caused by the different probabilities of the scattering direction. The scattering of the stellar light towards the observer is more probable for the inclined than for the face-on disk. Secondly, the scattered light of the outburst is first visible in the light curve of the inclined disk. This shows the importance of the light traveling time toward the observer. Since the lower part of the disk in the images is closer to the observer, the outburst will be seen there first. This is also visible in the first image of Fig.~\ref{fig:2D_burst_sca}. Furthermore, the outburst is stretched out compared to the face-on case, because parts of the disk with the same radial distance to the source will have different distances to the observer. This can be seen in the second of the scattering images. The third effect is a long afterglow caused by scattered photon packages with long distances to the observer. This wave of dim, scattered photon packages can be found in the upper half of the third scattering image.

\begin{figure}[!ht]
    \centering
    \includegraphics[width=1.\linewidth]{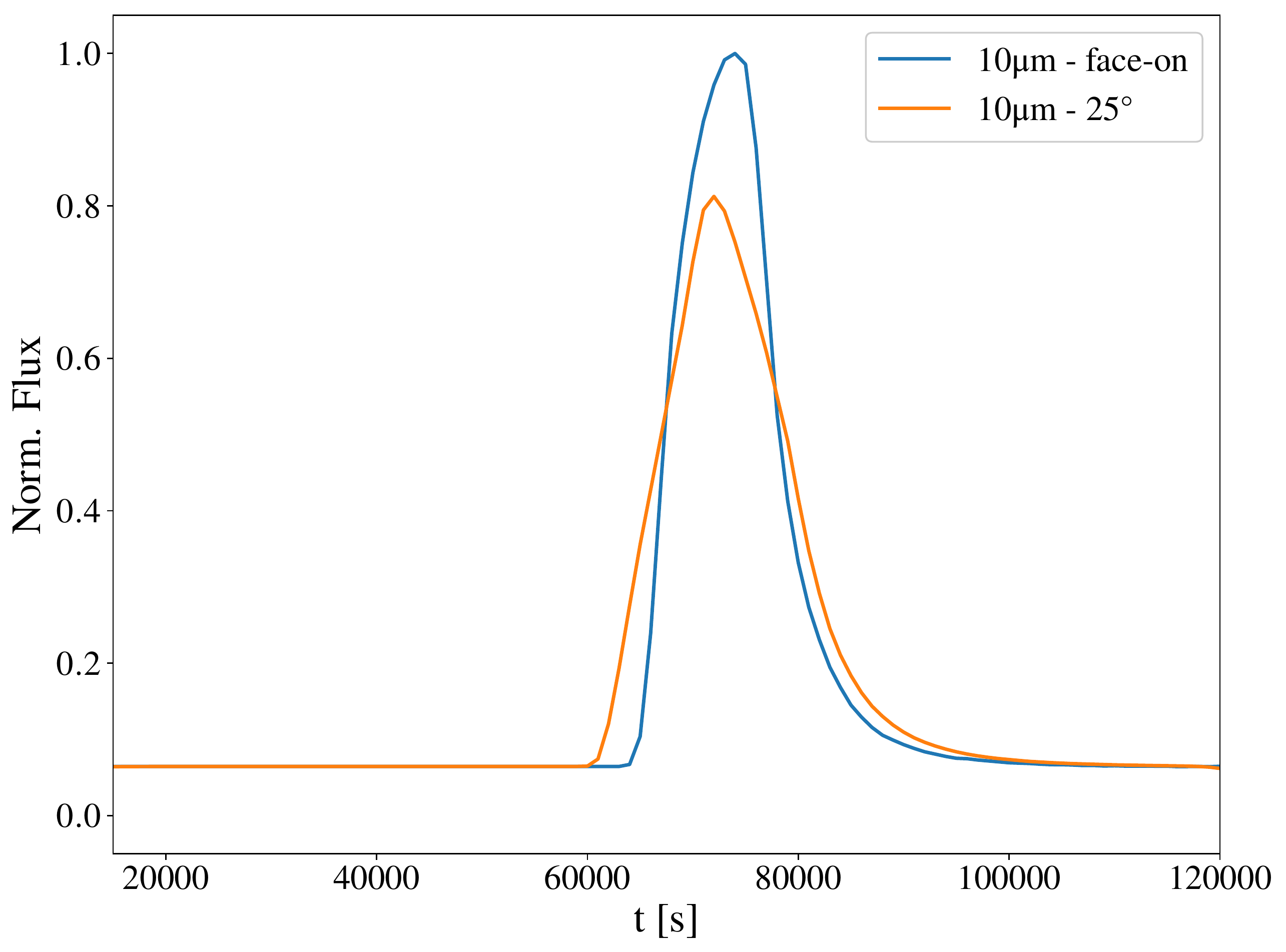}
    \caption{Light curve of the thermal reemission radiation (10\;$\upmu$m) of the outburst model discussed in Sect.~\ref{subsec:2D_burst}. The duration of the outburst is 10\,000\,s; the total simulation time is 120\,000\,s. During the outburst, the stellar luminosity is increased by a factor of 4. The flux is normalized with respect to the maximum of the integrated flux of all images of the different time steps.}
    \label{fig:burst_emi}
\end{figure}

\begin{figure*}[!h]
    \centering
    \includegraphics[width=1.\linewidth]{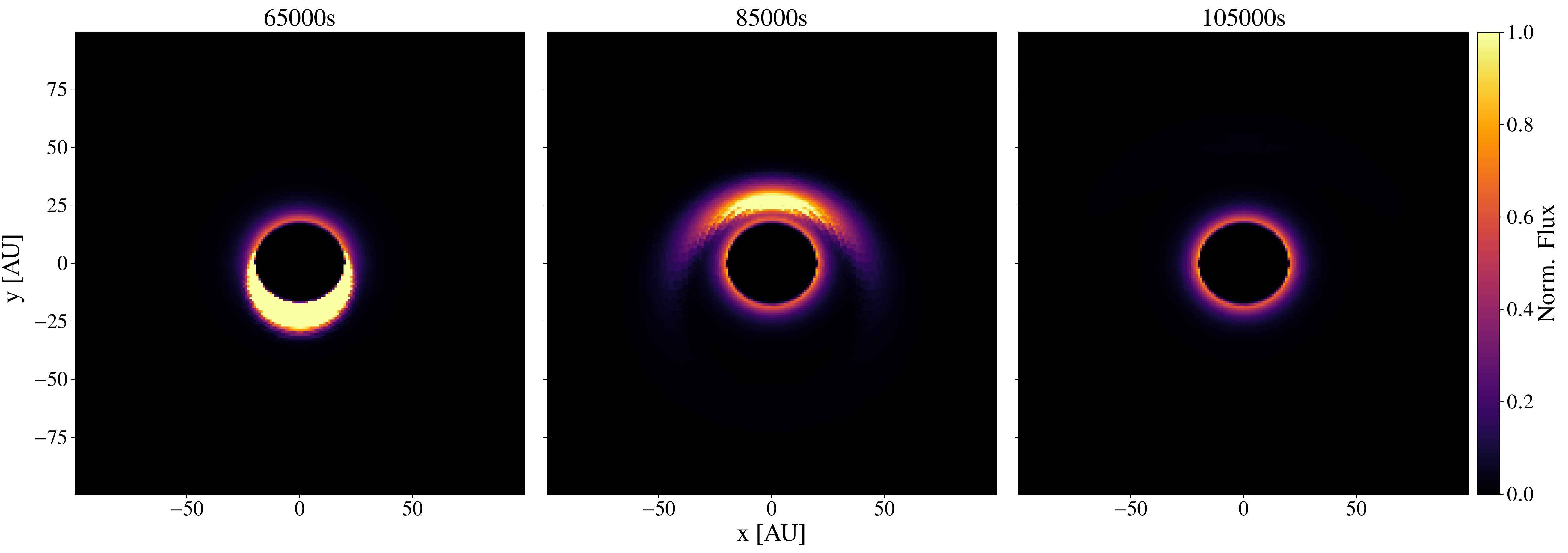}
    \caption{Images of the thermal reemission radiation of the outburst model after 65\,000, 85\,000 and 105\,000\,s for 10\;$\upmu$m described in Sect.~\ref{subsec:2D_burst} (temporal step width: 1000\,s;  $10^8$ photon packages per time step). The flux is normalized with respect to the maximum flux of all images of the different time steps.}
    \label{fig:2D_burst_emi}
\end{figure*}

The light curve of the simulation of the thermal reemission radiation is shown in Fig.~\ref{fig:burst_emi}. The corresponding images of the reemission radiation are given in Fig.~\ref{fig:2D_burst_emi}.
The main effect visible in the shape of the light curves is the light traveling time of the geometrical depth structure of the disk. While the heating and cooling time scales are short compared to the temporal step width, the heating of each layer of the disk in the reemission light curve will be visible with an offset relative to the light traveling time within the disk. Combined with the different light travel times due to the inclination of the disk, the light curve of the inclined disk shows a broader peak than the face-on light curve. It should be noted that the duration of the outburst relatively to the light travel times within the grid is a crucial factor for the shape of the emission peak, both in the case of the thermal reemission radiation and for the scattered radiation. There is also an afterglow visible in both light curves. However, the duration of the afterglow in the thermal reemission case is small compared to the one seen in the light curves of the scattered light. The reason for this is that the relative cold outer parts of the disk are not significantly contributing to the measured reemission radiation flux before or during the outburst. This also causes the emission peaks of the inclined and face-on disk to be closer together than in the case of scattering.

\subsection{2D disk outburst - High optical depth}\label{subsec:2D_burst_depth}
\begin{figure}[!ht]
    \centering
    \includegraphics[width=1.\linewidth]{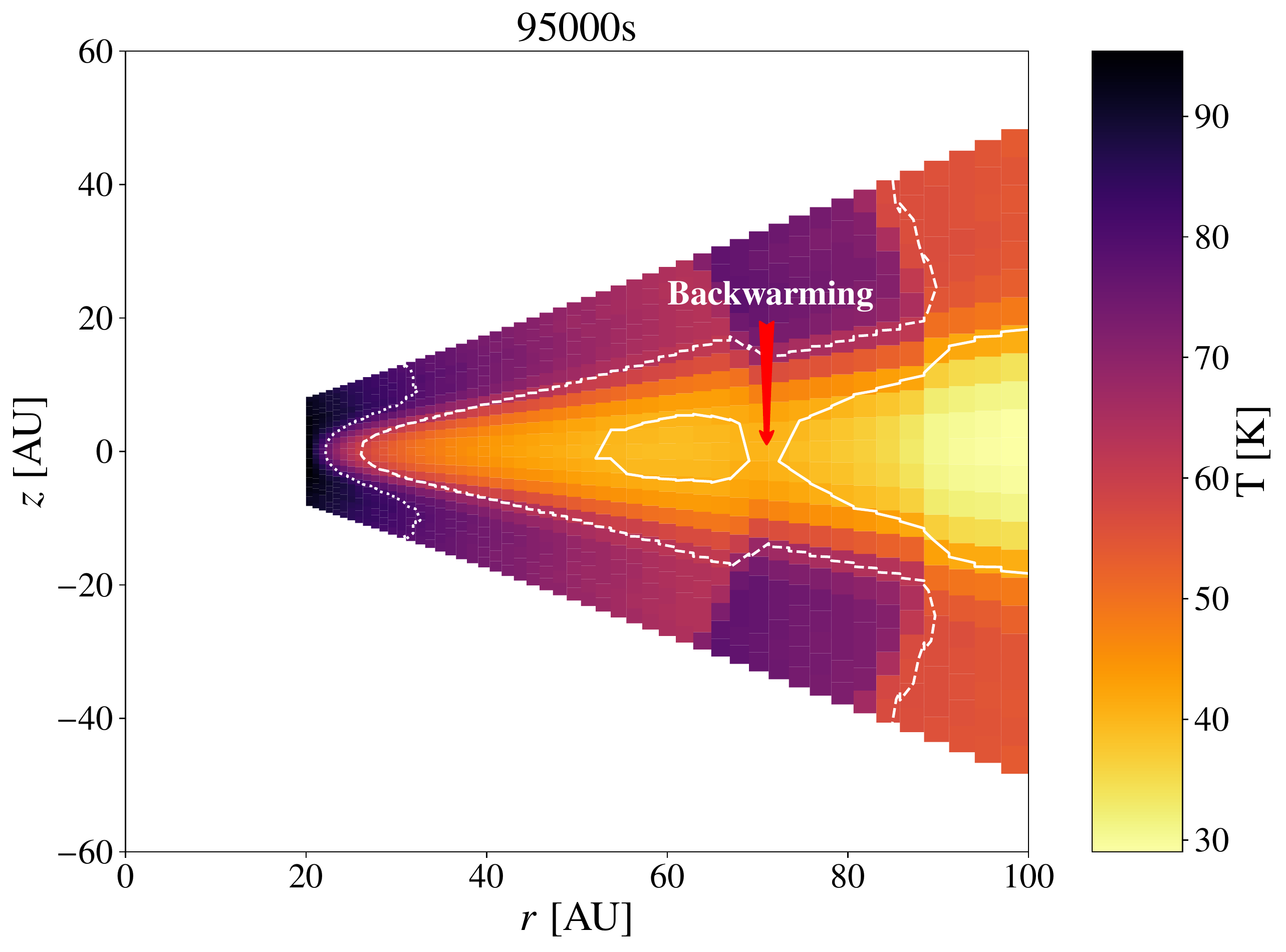}
    \caption{Vertical cut through the temperature distribution of the outburst disk model with high optical depth after 95\,000\,s described in Sect.~\ref{subsec:2D_burst_depth}.The temporal step width of the simulation is set to 4\,s; $10^5$ photon package are emitted per time step. The contour lines mark three levels of temperatures. The red arrow outlines a region in which the backwarming effect can be seen.}
    \label{fig:burst_back}
\end{figure}

To test the treatment of backwarming, that is the heating of regions of the disk by the thermal reemission radiation of different regions, we repeated the time-dependent temperature simulations of the disk model used in Sect.~\ref{subsec:2D_burst} with an optical depth of $\tau_V = 100$ in the midplane. In this case, the outer parts of the midplane of the disk are shielded from the direct emission of the star and can thus only be heated by the thermal reemission of the surrounding dust. A vertical cut through the temperature distribution of the high optical depth disk model after 95\,000\,s during the outburst is shown in Fig.~\ref{fig:burst_back}.
It can be seen that there is a warm region in the midplane around 70\,AU that is enclosed by colder regions. This means it was not heated by the stellar emission of the outburst, but by the thermal reemission of the dust above and below the midplane. It should also be noted, that the width and position of the warm spot do not correspond directly to the outburst in the optically thinner layers of the disk. This is caused by the delay between the heating by the stellar emission and the thermal reemission of the dust, as well as the density structure of the disk.

\section{Discussion and conclusions}  \label{sec:disc}
We presented an implementation of an algorithm for full time-dependent 3D Monte Carlo radiative transfer for the MCRT code \texttt{POLARIS}. It includes time-dependent methods for the simulation of temperature distributions and thermal reemission as well as scattering images. The method for the simulation of the temperature distribution focuses for the first time on the thermal properties of dust instead of gas. It can be used for the simulation of the radiative transfer of various astrophysical objects.

We discussed the influence of the newly introduced parameters temporal step width $\Delta t$ and number of photon packages per time step $N_{\rm ph}$. It was found that the temporal step width has to be small compared to the shortest cooling time of the systems, whilst the number of photon packages per time step has to be large enough for correct scattering and absorption treatment. Since all photon packages in the grid have to be loaded from and stored on a stack in each time step, the access speed and size of the CPU cache as well as the RAM are crucial parameters for the total computation time. Even though the treatment of the single photon packages in \texttt{POLARIS} is parallelized, the computation time is thus not necessarily decreasing linear with the number of threats. Temporal step width and number of photon packages per time step should therefore be chosen accordingly.

The time-dependent simulation of temperature distributions was successfully tested by simulating 1D and 2D models with different optical depth ($\tau_{\rm V}=0.01, 0.1, 1, 10, 100$). A deviation from the reference temperature distributions calculated with the stationary approach of \texttt{POLARIS} was only seen for the case of $\tau_{\rm V}=10$. The main reason for this is a slight inconsistency between the applied optical and calorimetric properties of the dust. While the optical properties are derived from measurements of crystalline olivine and fitted for astronomical purposes (see,~\citealp{Laor1993}), the thermal properties are derived from measurements of the heat capacity of basalt glass, obsidian glass and $\textrm{SiO}_2$ glass (see,~\citealp{DraineLi2001}). A discrepancy between optical and thermal properties leads to an underestimation of the temperature, if the heat capacity is overestimated compared to the optical properties. Better fitting dust data could therefore improve our method.

Finally, we tested a possible application of the code by simulating an outburst of a star with a circumstellar disk. Here, the temperature distribution, scattering images and light curve as well as thermal reemission images and light curve were calculated. The scattering and emission images and light curves were simulated for a face-on disk as well as for a disk with an inclination of $25^\circ$. The time-dependent ray tracing as well as the time-dependent scattering simulation are fast and can be performed independently of the simulation of the temperature distribution.

The effect of the heating of the disk by reemission radiation of the dust was shown by repeating the temperature simulation of the outburst model with an optical depth of $\tau_V = 100$ in the midplane.

It should be noted, that the presented method for the simulation of temperature distributions is most efficient at time scales from hours to weeks. To simulate variable events in circumstellar systems on longer time scales, it is needed to include the physics of the thermal coupling of the gas and dust phase. While this is out of the scope of this paper, the presented methods provide a valuable basis on which additional physical effects can be implemented. Since the methods for the time-dependent simulation of scattering and thermal reemission images are in principle independent of the temperature calculation, they can already be used on larger time scales. In the case of the thermal reemission simulations, the corresponding time-dependent temperature distributions could also be provided from hydrodynamics simulations.

Even though the presented method for the simulation of temperature distributions is in principle not dependent on the structure of the used dust distribution model, the computational afford increases with its complexity and optical depth. In order to represent the underlying optical depth structure for models with a high optical depth, an increased number of computational grid cells is needed. As discussed in Sect.~\ref{subsec:nph} a high number of grid cells requires a higher number of photon packages per time step. Steep gradients in the dust density structure will also lead to high temperature gradients that require smaller temporal step widths. For example, in the case of a circumstellar disk with optical depth $\tau_V > 10^5$ the temperature simulation would require at minimum $10^4$ photon packages per time step with a temporal step with of around $0.01\,$s. Depending on the radial extent of the model, the total number of photon packages processed per time step are then about $10^8$ to $10^9$. This corresponds roughly to the computational afford of a single stationary radiative transfer simulation and should therefore be considered when setting up time-dependent simulations.

Future applications are aiming at the simulation of images and light curves of observed variable illumination and heating sources embedded in dust distributions, such as young stellar objects embedded in their parental molecular cloud or surrounded by circumstellar disks, to study their geometrical structure and dust properties. Potential improvements of the numerical implementation cover a faster memory management as well as the possibility to start the temperature simulation from a steady state instead of including the simulation of the initial heating process of the disk.

\section*{ORCID iDs}
A. Bensberg \orcidlink{00000-0001-6789-0296}
\href{https://orcid.org/0000-0001-6789-0296}
     {https://orcid.org/0000-0001-6789-0296}\\
S. Wolf \orcidlink{0000-0001-7841-3452}
\href{https://orcid.org/0000-0001-7841-3452}
     {https://orcid.org/0000-0001-7841-3452} 


\bibliographystyle{aa} 
\bibliography{lit} 

\begin{appendix}
\section{Absorption estimator}\label{sec:abs_est}
Following the discussion of the probability density functions of absorption optical depth $\tau_{\rm abs}$ and interaction optical depth $\tau_{\rm ext}$ in~\citeauthor{Lucy1999}~(\citeyear{Lucy1999}; further discussed in~\citealp{Krieger2020}), we are subsequently focusing on the different ways to treat the absorption of a photon package without explicitly taking scattering into account.
We consider the energy of a photon package deposited while passing an optical depth interval of $[\tau_1,\tau_2)$. The absorption probability for this interval can be written as $\int_{\tau_1}^{\tau_2} \textrm{d}\tau e^{-\tau}$. For simplicity, we are normalizing the energy of a photon package to 1. It will be shown, that the same energy is deposited in all three approaches to handle absorption:
\paragraph{Case 1: No continuous absorption, only single absorption events}
In the case of only single absorption events, all energy is deposited in the grid and the photon package is deleted after the event. The energy $\varepsilon$ deposited in the grid is then given by:
\begin{equation}
    \varepsilon = \int_{\tau_1}^{\tau_2} \textrm{d}\tau e^{-\tau} \cdot 1 = e^{-\tau_1} - e^{-\tau_2}.
\end{equation}
\paragraph{Case 2: Continuous absorption, photon package is loosing energy}
We are now considering a photon package that is continuously loosing energy which is deposited in the grid, that is $(e^{-\tau_1} - e^{-\tau})$. In the case of an absorption event, the photon package leaves all remaining energy in the grid and is deleted. 
The deposited energy can then be calculated with:
\begin{equation}
    \begin{aligned}
        \varepsilon = &\int_{\tau_1}^{\tau_2} \textrm{d}\tau e^{-\tau} \left( (e^{-\tau_1} - e^{-\tau}) + (1-e^{-\tau})\right)\\
        &+ \int_{\tau_2}^{\infty} \textrm{d}\tau e^{-\tau} (e^{-\tau_1} - e^{-\tau_2}) = e^{-\tau_1} - e^{-\tau_2}. 
    \end{aligned}
\end{equation}
\paragraph{Case 3: Continuous absorption, photon energy stays constant}
The last case is similar to case 2, except that the energy of the photon package stays constant and no energy is deposited in the grid when the photon is deleted. Thus, a photon package traveling a distance $\delta \tau$ leaves the energy $\delta \tau$ in the grid:
\begin{equation}
    \varepsilon = \int_{\tau_1}^{\tau_2} \textrm{d}\tau e^{-\tau} (\tau-\tau_1) + \int_{\tau_2}^{\infty} \textrm{d}\tau e^{-\tau} (\tau_2 - \tau_1) = e^{-\tau_1} - e^{-\tau_2}.
\end{equation}

It can be seen that the deposited energy is the same in all three cases. However, in a numerical simulation, we do not always reach this limit. In the simulations presented in this paper, we chose case 3, in which the energy is distributed evenly in the grid even for small numbers of photon packages.

\section{Exemplary study: Dust shell outburst model}\label{sec:sn_echo}
In Sect.~\ref{subsec:timedep_ray} we presented a time-dependent ray tracing method for the simulation of thermal reemission images. A simple test for the correct treatment of light travel times is an outburst of a central source embedded in a spherical dust distribution. We thus set up a one-dimensional spherical model containing five dust shells with constant dust density (inner radius $R_{\rm in} = 0.1$\,AU, outer radius $R_{\rm out} = 10$\,AU, optical depth $\tau_{\rm V}=55$) and an initial dust temperature of 50\,K. An overview of all model parameters can be found in Tab.~\ref{tab:sn_echo}.
\begin{table}[!h]
  \begin{center}
    \caption{Overview of the parameter of the spherical 1D dust shell model discussed in Sect.~\ref{sec:sn_echo}.}
    \label{tab:sn_echo}
    \begin{tabular}{llc}
        \hline
        \hline
        \rule{0pt}{2ex}
        \textbf{Parameter} & & \textbf{Value}\\
        \hline
        \rule{0pt}{3ex}
        Inner radius & $R_{\rm{in}}[$AU$]$ & 0.1 \\
        \rule{0pt}{1ex}
        Outer radius & $R_{\rm{out}}[$AU$]$ & 10 \\
        \rule{0pt}{1ex}
        Total dust mass & $M_\text{dust}$[M$_\odot$] & $10^{-7}$\\
        \rule{0pt}{1ex}
        Optical depth & $\tau_{\rm V}$ & 55\\
        \rule{0pt}{1ex}
        Number of cells in & &\\
        \rule{0pt}{1ex}
        ... radial direction & $N_{r}$ &  $5$\\
        \rule{0pt}{1ex}
        ... azimuthal direction & $N_{\phi}$ & $1$\\
        \rule{0pt}{1ex}
        ... polar direction & $N_{\theta}$ &  $1$\\
        \hline
    \end{tabular}
  \end{center}
\end{table}

For the sake of simplicity, we model a heatwave by raising the dust temperature of every shell to 75\,K, beginning from the innermost shell according to the light travel time through the entire model. After each time step, the dust temperature is reset to 50\,K and the temperature of the next cell is set to 75\,K. The temporal step width of $\Delta t = 1000\,$s is chosen such that one time step corresponds to the light traveling time through one of the dust shells in radial direction.

\begin{figure}[!hb]
    \centering
    \includegraphics[width=.95\linewidth]{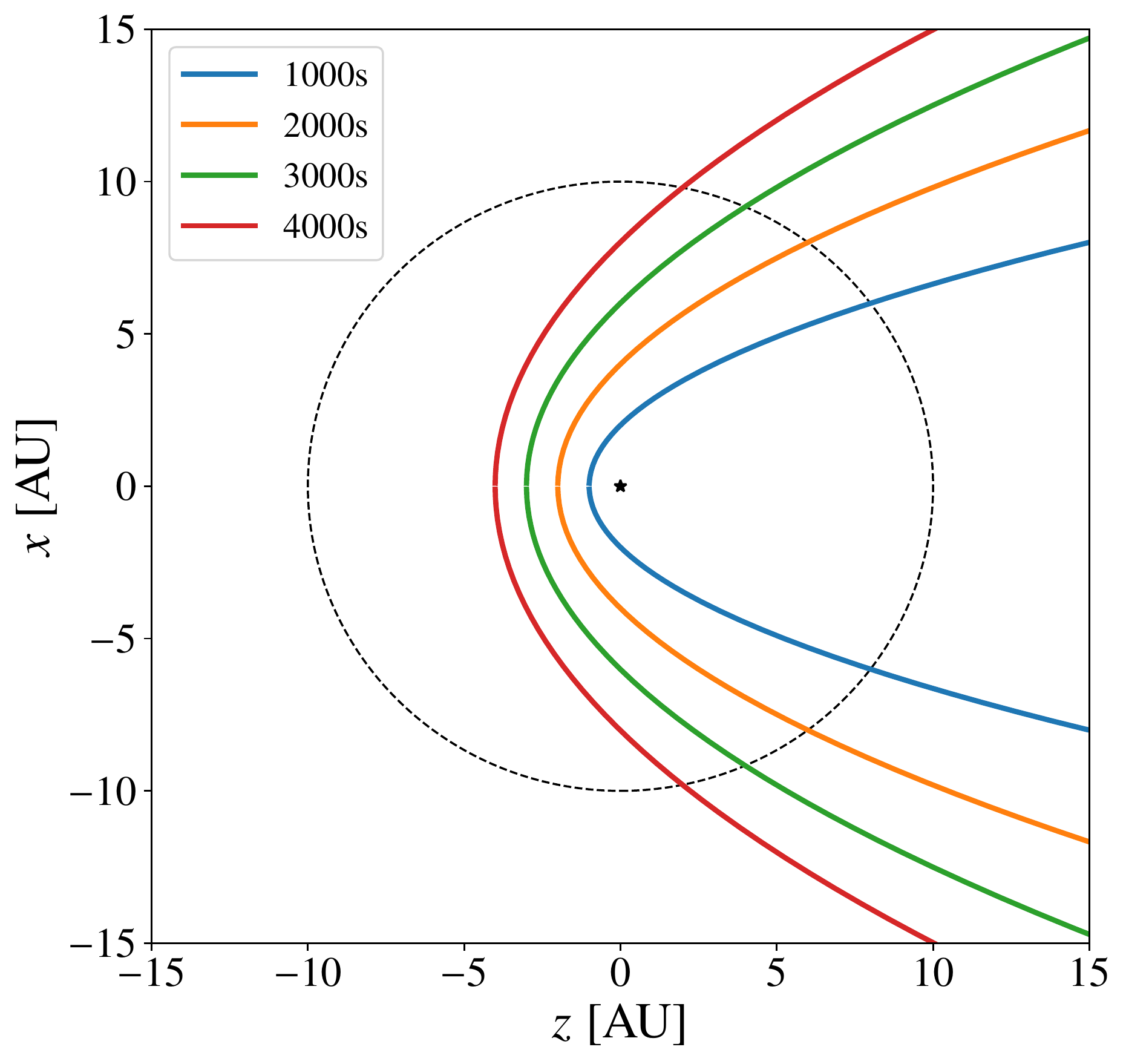}
    \caption{Dust shell outburst model: Regions of dust reemission radiation seen by an observer for different time steps after an outburst for the model described in Sect.~\ref{sec:sn_echo} calculated using Eq.~\ref{eq:sn_echo}. The dashed line indicates the outer radius of the dust distribution. The position of the central source is indicated by an asterisk.}
    \label{fig:sn_echo}
\end{figure}

The position of the reemitting, heated dust as seen from the observer at a time $t$ after which the outburst is first detected can be described analytically by a parabola known for example from supernova light echoes (see,~\citealp{Wright1980}):
\begin{equation}\label{eq:sn_echo}
    z = \frac{x^2}{2ct} - \frac{1}{2}ct,
\end{equation}
where $z$ is the distance to the observer, $x$ the distance parallel to the observering plane, and $c$ the speed of light. The corresponding parabolas for the outburst model described above are shown in Fig.~\ref{fig:sn_echo}.

\begin{figure*}[!ht]
    \centering
    \includegraphics[width=1.\linewidth]{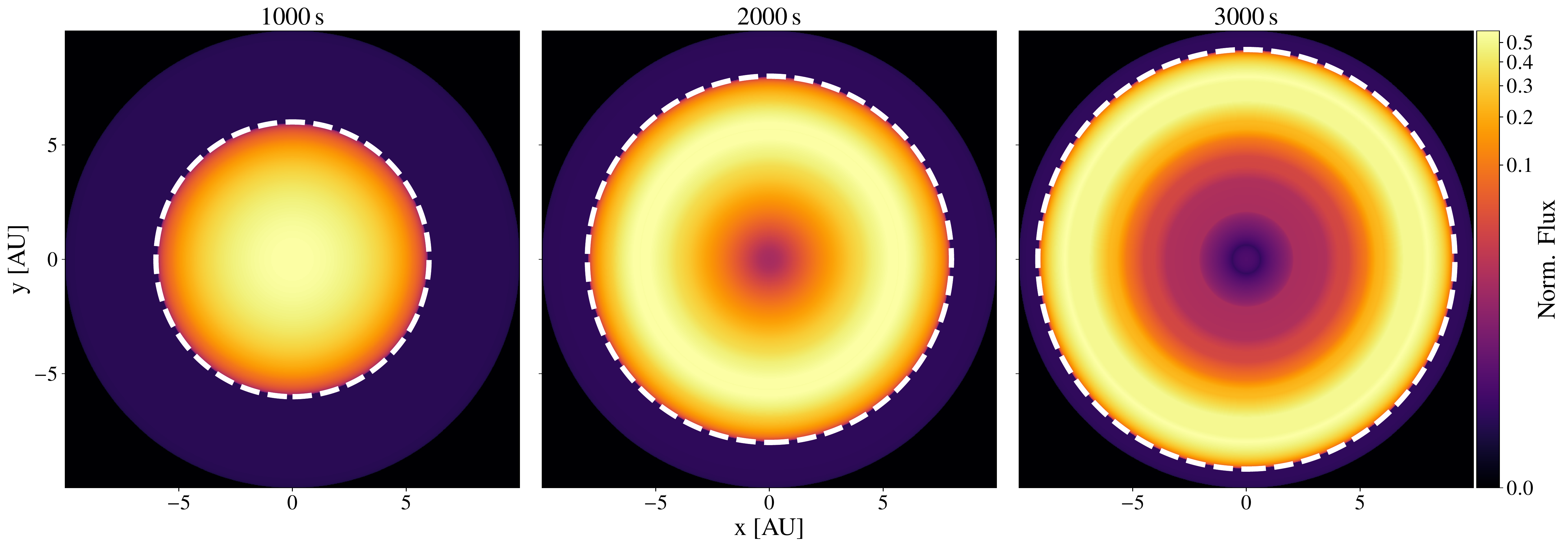}
    \caption{Images of the thermal reemission radiation of the dust shell outburst model after 1000, 2000 and 3000\,s at a wavelength of 10\;$\upmu$m as described in Sect.~\ref{sec:sn_echo} (temporal step width: 1000\,s). The flux is normalized with respect to the maximum flux of the image after 1\,000\,s. The white dotted lines indicate the maximum radius of the parabola depicted in Fig.~\ref{fig:sn_echo}.}
    \label{fig:sn_echo_1x3}
\end{figure*}
These analytic predictions can now be compared to the images of the thermal reemission radiation obtained using the time-dependent ray tracing algorithm described in Sect.~\ref{subsec:timedep_ray}. The thermal reemission images for three time steps after the first detection of the outburst can be found in Fig.~\ref{fig:sn_echo_1x3}. The contour lines represent the maximum radius of the parabolas shown in Fig.~\ref{fig:sn_echo}. 
The resulting images obtained with the time-dependent ray tracing algorithm match the analytic predictions perfectly.
\end{appendix}
\end{document}